\documentclass[]{interact}

\usepackage{epstopdf}
\usepackage[caption=false]{subfig}

\usepackage[longnamesfirst,sort]{natbib}
\bibpunct[, ]{(}{)}{;}{a}{,}{,}
\renewcommand\bibfont{\fontsize{10}{12}\selectfont}
\usepackage[utf8]{inputenc}
\usepackage{amsmath}
\usepackage{amsfonts}
\usepackage{algpseudocode}
\usepackage{graphicx}
\usepackage{varwidth} 
\usepackage{url}
\usepackage[ruled]{algorithm2e}

\begin{document}

\title{ChatGPT: More than a “Weapon of Mass Deception”\\
Ethical challenges and responses from the Human-Centered Artificial Intelligence (HCAI) perspective}

\author{\name{Alejo José G. Sison\textsuperscript{1}, Marco Tulio Daza\thanks{Author emails: ajsison@unav.es, mdazaramire@alumni.unav.es, 201905616@alu.comillas.edu, and ecgarrido@icade.comillas.edu}\textsuperscript{1,2}, Roberto Gozalo-Brizuela\textsuperscript{3} and Eduardo C. Garrido-Merchán\textsuperscript{3}}
\affil{ \textsuperscript{1} School of Economics and Business Administration, Institute of Data Science and Artificial Intelligence (DATAI), Universidad de Navarra, Pamplona, Spain\\ \textsuperscript{2} Information Systems Department, University Center for Economic and Administrative Sciences (CUCEA), University of Guadalajara, Guadalajara, Mexico \\ \textsuperscript{3} Quantitative Methods Department, Universidad Pontificia Comillas, Madrid, Spain}
}

\maketitle

\begin{abstract}
This article explores the ethical problems arising from the use of ChatGPT as a kind of generative AI and suggests responses based on the Human-Centered Artificial Intelligence (HCAI) framework. The HCAI framework is appropriate because it understands technology above all as a tool to empower, augment, and enhance human agency while referring to human wellbeing as a “grand challenge”, thus perfectly aligning itself with ethics, the science of human flourishing. Further, HCAI provides objectives, principles, procedures, and structures for reliable, safe, and trustworthy AI which we apply to our ChatGPT assessments. The main danger ChatGPT presents is the propensity to be used as a “weapon of mass deception” (WMD) and an enabler of criminal activities involving deceit. We review technical specifications to better comprehend its potentials and limitations. We then suggest both technical (watermarking, styleme, detectors, and fact-checkers) and non-technical measures (terms of use, transparency, educator considerations, HITL) to mitigate ChatGPT misuse or abuse and recommend best uses (creative writing, non-creative writing, teaching and learning). We conclude with considerations regarding the role of humans in ensuring the proper use of ChatGPT for individual and social wellbeing.

\end{abstract}

Keywords: ChatGPT, generative AI, HCAI, combating disinformation, AI ethics
\\

\textit{“The most important outcome of that journey is that it will compel us to understand what it means to be human” (Gary Marcus)} 

\textit{"I do really believe that creativity is computational... It is something we understand the principles behind. So, it’s only a matter of having…neural nets or models that are smarter." (Yoshua Bengio, Turing award.)}
   
\section{Introduction}

In a review of large Generative AI models, \cite {gozalo2023chatgpt} list among the work’s limitations the “lack of an understanding of ethics” which prevents the new technology from achieving full potential. This paper seeks to rise to this challenge. 

Since its rollout in November 2022, a host of ethical issues has arisen from the use of ChatGPT: bias, privacy, misinformation, and job displacement, among others. But this did not dampen public enthusiasm. By February 2023, ChatGPT broke the record for application user-base growth, rising to 100M in just two months \citep{LWAI205115} and reaching a market value close to 30 billion USD \citep{Roose2023race90}. 

In 2022, venture capitalists invested 2.7 billion USD in 110 generative-AI startups \citep{Economist202381}. In consequence, alternatives to ChatGPT did not take long in appearing: Deep Mind’s Sparrow, Google’s LaMDA/Bard, Antrophic’s Claude, and an open-source transformer by Stability AI \citep{Roose2023race90, Romero2023m17, Konrad2023break, Metz2023116, Roose2023119}. 

The magnitude of investments is remarkable, considering that initially, there was no established business model for ChatGPT \citep{Vincent202382}. A subscription model, 'ChatGPT Plus', offering greater availability, faster responses, and priority access to new features for 20 USD/month is clearly inadequate \citep{Schreckinger202384}. Hence the race to team up with Microsoft, combining ChatGPT with Bing (search engine) and Edge (browser). The experiment ended up largely a disaster, not only because of ChatGPT’s unpredictable, erroneous, and creepy responses, but also because a Large Language Model (LLM) is much costlier to run than a search engine, while admitting fewer ads, and overall less traffic \citep{Romero2023124}. Google’s hasty attempt to compete with its own LLM Bard turned out no better, erasing 100 billion USD of market value \citep{Marcus2023121}. As \cite{WarzelVertigo2023} commented, “The idea of generative AI as a new frontier for accessing knowledge, streamlining busywork, and assisting the creative process might exhilarate you. It should also unnerve you. If you’re cynical about technology (and you have every reason to be), it will terrify you”. LLM chatbots only make search engine outputs unreliable.

Together with huge amounts of money came the hype about what ChatGPT can do: dating help, crime fighting, psychotherapy, and so forth \citep{Romero202270, Khullar2023158}. For many, ChatGPT was only or above all a means to earn money, gain social media presence, or (somewhat dubiously) take a step toward singularity \citep{Romero202357}. 

Besides investigating the ethical issues involving ChatGPT, we need a better understanding of its technical specifications, both potentials and limitations, to ascertain mitigation measures for misuse or abuse, with a view to enhancing human flourishing or wellbeing. This is the first among the six grand challenges, together with responsible design, respect for privacy, human-centered design, appropriate governance and oversight, and respect for human cognitive capacities recently identified for a human-centered AI (HCAI) \citep{Garibay2023}, the theoretical lens through which we carry out evaluations and assessments. Pioneered by \cite{Shneiderman2020AIS, Shneiderman2020IEEE, Shneiderman2020IJHCI, Shneiderman2022HCAI, Shneiderman2022medium, Shneiderman2021ACM}, HCAI contends that human control is compatible with a high degree of automation, that AI works best not when emulating humans but when empowering them, and that proper AI governance should aim, above all, in making AI reliable, safe, and trustworthy. 

This paper is structured as follows. Section II presents the ethical challenges posed by ChatGPT as a kind of generative AI, focusing on its employment as a “weapon of mass deception” (WMD). Section III deals with technical potentials and limitations of ChatGPT with a view to the ethical challenges. Section IV suggests both technical and non-technical measures to mitigate ChatGPT misuse, as well as recommendations for best uses. And section V concludes, with considerations about how ChatGPT can affect human wellbeing. 

But before proceeding, a brief statement on limitations. The first refers to ChatGPT’s recency and rapid yet unsettled development which make us rely heavily on specialized news and technical blog posts, given the dearth of peer-reviewed journal articles. We are aware that being among the first movers in this field comes with a price. Secondly, despite the abundant literature on AI ethics, we wanted to narrow our focus on ChatGPT, its specific problems and recommendations for engagement. Of course issues such as bias, privacy, job displacement, and manipulation cut across the whole range of AI, yet our intention was to concentrate on how these challenges are manifested in ChatGPT, given its purportedly revolutionary nature. And thirdly, in consonance with the general theme of human-computer interaction, we wish to address ethical issues more broadly, integrating both technical and non-technical aspects, albeit toward a clearly human or social goal of wellbeing in accord with the HCAI framework. Business references underscore the importance of economic motives in decisions regarding technology development and deployment. Our goal has been to write a paper that is “technical enough” and “humanistic enough” to deal appropriately and adequately with the ethical challenges and responses surrounding ChatGPT, without falling into either extreme. Although a lot more can be said both in the technical and ethical dimensions, however, we perceive great value in carving out and staking claim to a common ground between the two.

\section{The ethical challenges posed by ChatGPT as a kind of generative AI}

Generative AI models are characterized by the ability to produce new content and are found in a variety of fields \citep {gozalo2023chatgpt}. Outputs include text (ChatGPT, Claude, and Bard), code (GitHub Copilot and Codex), images (Midjourney, DALL-E2, and Stable Diffusion), music (MuseNet and MusicLM), video (Synthesia and Elai), and even human-like voices (VALL-E). 
The deployment of generative AI raises several ethical concerns \citep{Huang2023109}. A recent survey \citep{Daza2022} groups ethical issues concerning AI in business into five categories, all of which are applicable to generative AI:
\begin{enumerate}
    \item Foundational issues: Do LLMs possess sentience? \citep{Tiku2022} Are they a step closer to AGI? \citep{Liang2023122, Marchese202261, Marcus2023108, Marcus2023125, SisonRedin2021}. 
    \item Privacy: Generative AI training data may infringe on privacy and violate copyrights \citep{Goldman202344, Vincent2023legal, Dixit2023legal, Setty202343, Wiggers202380, Growcoot2023141}. Deepfake porn \citep{Romero2023m17, Gorrell2023}.
    \item Bias: Generative AI can reproduce and amplify biases \citep{Kriebitz2020, Marcus2023125}. 
    \item Employment and automation: Generative AI can increase job displacement \citep{Lowrey2023}.
    \item Social media and public discourse: Generative AI creates echo-chambers \citep{Levy2021}and produces emotional contagion \citep{Kramer2014}. It can also exploit psychological vulnerabilities and become a tool for manipulation \citep{Parker2017}.
\end{enumerate}

The different ways to engage with these issues are well known. Behind the foundational issues of sentience and AGI is anthropomorphism, the tendency to project human agency onto things for a semblance of understanding (e.g., “the doll smiles because it likes me”). In the case of machines, it is called the “Eliza effect”, named after a 1960s chatbot \citep{Weizenbaum1966}. Developers have to be transparent about how their models are designed and contribute to increasing public AI literacy. Privacy concerns can be addressed through “privacy by design” \citep{Cavoukian2009} and practices such as data minimization, encryption, and informed consent. Legal disputes regarding copyrighted materials can be avoided if companies adopt the three C's: credit, compensation, and consent \citep{Clark202323}. Companies should also adhere to the doctrine of "fair use," which permits reproducing copyrighted materials in certain circumstances \citep{Wiggers202380}. 

To mitigate bias, it's important to use diverse and representative training data \citep{Kusner2020}. Techniques such as "fairness through unawareness" can be used to remove sensitive information from inputs. Working with a diverse, interdisciplinary team and promoting stakeholder engagement leads to less bias, just as algorithmic audits can help prevent hallucinations. To deal with job displacement, policies and programs to support workers through retraining and upskilling can be developed. And for the negative impacts of social media on public health and democracy, robust laws, frameworks and guidelines for responsible AI use in content creation and dissemination can be implemented.

Let us now turn to ChatGPT.

\subsection{ChatGPT as a "weapon of mass deception" (WMD)}

The Internet's game-changing power is shown, among others, in driving down the costs of reproducing and distributing digital content. Similarly, ChatGPT has lowered the marginal costs of producing new and original human-sounding texts to practically zero (once the costs of construction, training, maintenance, and so forth are covered) \citep{Klein20238}. Further, the model is very user friendly (user experience or UX), responding to natural language prompts, hardly requiring any user training (user interface or UI). ChatGPT presents a ready-to-use uniquely synthesized text, not links to references like search engines. It can even be prompted to follow a particular language style of a period or an author, “mimicking creativity” \citep{Thompson20221}. Because ChatGPT produces highly coherent, natural-sounding, and human-like responses, users find them convincing and readily trust them, even if inaccurate. The “automation bias”, occurring when humans blindly accept machine responses as correct, without verifying or even disregarding contradictory information is extensively documented. After all, machines are objective, do not grow tired or get emotional; they have instant access to ever greater information \citep{Metz202330}.

However, there are a few ethical issues particularly relevant to ChatGPT. ChatGPT has been dubbed “the world’s best chatbot” \citep{Romero202232}. Even informed subjects would most likely need several interactions to determine they are engaged with a machine, in what amounts to ChatGPT passing the Turing Test with flying colors \citep{Metz202330}. Many people engage with ChatGPT for social reasons, wanting to have a conversation (friendly, romantic, therapeutic, and so forth) or for entertainment \citep{Weise2023144}, where accuracy and precision do not matter. As a machine, ChatGPT doesn’t really have a personality; it simply reflects the prompter’s desires for the sake of conversation \citep{Metz2023154}.

In February 2023, Microsoft hastily released a test version of its Bing search engine incorporating an OpenAI chatbot. Soon, a “shadow-self” named Sydney emerged \citep{Romero2023d20m2}: one who threatened users, confused dates, gaslit interlocutors, suffered existential breakdowns, and even professed love \citep{Roose2023d16m2}. Microsoft received flak for going against its own principles of fair, reliable, safe, and secure AI design \citep{Blackman2023151}. However, despite its creepiness, chatbot conversations could be addicting, precisely because models are trained to give the “desired” response. 

When asked to self-disclose, ChatGPT declares it is a mere LLM: “As a machine learning model, I do not have feelings, beliefs or consciousness. [...] I am different from humans in several ways when it comes to generating my responses. Some of the main differences include: lack of consciousness, limited understanding, lack of context, limited creativity, and limited ability to reason.”.

In this respect, it does not deceive or lie; since phenomenal consciousness or meta-cognition is not necessarily related with any form of intelligence \citep{merchan2022independence}, but placed in the wrong hands, ChatGPT can certainly be used as a “weapon of mass deception” (WMD). As Gordon Crovitz of NewsGuard (a company which tracks disinformation online) affirmed, “This tool is going to be the most powerful tool for spreading misinformation that has ever been on the internet. (...) Crafting a false narrative can now be done at dramatic scale, and much more frequently –it’s like having AI agents contributing to disinformation” \citep{Hsu2023120}. 

The main ethical issue concerning ChatGPT is its use as a tool for deception. ChatGPT and similar models generate responses that are highly believable, because of its mostly impeccable syntax and language, although not entirely true, occasionally mixing fact with fiction (“hallucinations”) \citep{Bashir20223, Ferus20236}. A less polite term for these “hallucinations” is “bullshit”, in the words of American philosopher Harry G. Frankfurt. As Wharton professor Mollick explains, “bullshit is convincing sounding nonsense, devoid of truth, and AI is very good at creating it. You can ask it to describe how we know dinosaurs had a civilization, and it will happily make up a whole set of facts explaining” \citep{Terwiesch202363}. The purpose of “bullshit” is not to inform but to deceive. Yet despite generating “bullshit”, ChatGPT is not legally responsible for its output \citep{Romero2023157}.

ChatGPT does not do this intentionally (it cannot  have intentions). In determining word sequence, it does not consider nor can it distinguish between truth and falsehood; it simply predicts the next word based on statistical correlations in the training data and the prompt as an advanced autocomplete function. Its training data set is not properly curated nor verified; and we don’t know sources. Unsurprisingly, ChatGPT has been found deficient in logic and in mathematical skills \citep{Ott2023, Roose202315, Marcus202319, Agostino202317}.

Above all, ChatGPT is not a search engine, designed to provide information, nor is it an Internet gateway \citep{Romero2023124}. Although search engines are rigid, they don’t invent things. ChatGPT is more intuitive and flexible, but unreliable \citep{Romero202232}. Some experts think search and text generation are completely different functions; and while conversational replies may be preferable to links, combining both in a hybrid model entails enormous challenges \citep{Romero202336}.

The deceptive potential of ChatGPT can be employed in at least three different ways: in the academe, by passing off work as one’s own or by attributing co-authorship; through disinformation campaigns (deepfake texts and impersonation) especially in social media; and by enabling criminal activities through malware production for phishing, unauthorized access and extorting ransom (encryption/decryption), and creating darkweb pages, all of which involve some form of deceit.

\subsubsection{Deception in academic work}

Firstly, a special class of deception concerns academic work and knowledge production \citep{Bowman20224}. Proper research consists of posing questions, gathering information, understanding it, deciding what’s important, and actually writing. ChatGPT is not intellectually curious nor does it pose research questions unprompted. It gathers a lot of information but indiscriminately, unable to separate truth from falsehood or fact from opinion, and only until a given date (it has no Internet connection and cannot “refresh” its data, although it can assimilate information from prompts). It does not understand data semantically, grasping meaning, since it only sees formal, statistical patterns between words \citep{Agostino202317}. Its decision to use one word instead of another is driven exclusively by statistics. ChatGPT acts like a blender or food processor that combines and regurgitates whatever information it is fed in a new or even “original” and human-like form, providing outputs it doesn’t understand while completely oblivious of purpose. It has also been called a “stochastic parrot”, randomly stitching words together without reference or meaning \citep{Bender2021}.

The issue is not so much using ChatGPT to save the trouble of gathering information and synthesizing it, but in passing off output as one’s own. This may not be a problem in preparing message templates and other routinary tasks \citep{Friedersdorf20235}, such as job applications and résumés \citep{Morrone202354}. But it would be unacceptable for an academic exercise. Why? Because the whole point is to learn the skills of information gathering, judgment, and synthesis, expressing thoughts in one’s words with fresh insights. Hardly any of this occurs when all one does is to enter a prompt. That’s why in this context, ChatGPT use is fraudulent. It’s not supposed to replace thinking or learning to think for oneself.

The unscrupulous may argue that ChatGPT has rendered these research skills obsolete, justifying the effort and time-saving measures. But even when putting deception aside, the majority of people would still disagree, for the same reasons we insist on school-children learning math skills, despite the ubiquity of 100\% accurate calculators (a degree of reliability ChatGPT cannot reach). Overdependence on technology has the perverse effect of deskilling which stymies the very capacity for innovation.

So the main ethical issue does not lie on whether the output is original or new (which it can very well be) or whether it is possible to plagiarize a machine output (doubtful), but that one misappropriates the work of “another”, misrepresenting it as one’s own, while failing to develop and demonstrate competence. On the one hand, there is no person to steal the output from, and on the other, ChatGPT does not only collect facts, but actually generates texts \citep{Barnett202371}. In any case, deception by falsely representing an academic work is a breach of honesty and integrity. It's copying, cheating, and lying.

A particular way of unethically passing off ChatGPT output takes place in exams. Reportedly, ChatGPT passes evidence and torts in the multiple-choice section of a multistate bar exam, although not the overall exam \citep{Bommarito2022}. In a related event, Joshua Browder, founder of DoNotPay, an organization promoting equity in legal assistance, planned to use GPT-J to assist defendants in traffic court, despite doubtful permissibility \citep{Schreckinger2023law}. In the end, Browder had to back down under threat of sanctions for bringing a “robot lawyer” into the court \citep{Romero2023m16}.

It has also been reported that ChatGPT would pass Operations Management in the Wharton MBA program \citep{Terwiesch202349}. It performs well with basic operations management and process analysis questions, including case studies, but fails with elementary math. In its favor, however, is a responsiveness to cues, modifying outputs accordingly. ChatGPT may pass certain school tests, but that is all it can do. We cannot infer that by passing tests, it can actually put into practice (as humans do) the knowledge, skills, aptitudes, and functions tests were supposed to verify \citep{Mitchell2023130}.

In medicine, ChatGPT is said to perform at or near passing for all three steps (step 1: basic science, pharmacology, and pathophysiology; step 2: clinical reasoning, medical management, and bioethics/ clinics; step 3: post-graduate medical education) of the U.S. Medical Licensing Exam \citep{Kung202260}. Study authors (which include ChatGPT!) affirm it has the potential to assist with medical education and clinical decision making. Yet surprisingly, no one suggests that ChatGPT receive a medical license, despite passing the exams.

Another deceptive practice is attributing co-authorship of an academic work to ChatGPT. In January 2023, it was listed as co-author in at least four medical articles \citep{Stokel2023}. What seemed at first as careful acknowledgment of intellectual debt actually rests on a cognitive error, because machines are tools or instruments, not authors. Only humans who can grasp the purpose of their work and willingly engage in research qualify as authors. Humans alone can accept the legal, scientific, moral and social responsibility for their publications or consent to terms of use and distribution. Experimental scientists do not attribute co-authorial rights to instruments or software; the same holds for ChatGPT. It is highly misleading to list ChatGPT as co-author as this implies false, unscientific claims and attributions \citep{Marcus202319}. The journal Nature, together with all Springer Nature Journals, establish two new author guidelines: “First, no LLM tool will be accepted as a credited author on a research paper. That is because any attribution of authorship carries with it accountability for the work, and AI tools cannot take such responsibility. Second, researchers using LLM tools should document this use in the methods or acknowledgment sections. If a paper does not include these sections, the introduction or another appropriate section can be used to document the use of the LLM.” \citep{nature2023guide}.

\subsubsection{Disinformation campaigns}

Secondly, outside the academic domain, ChatGPT can be used to deceive or mislead people (disinformation) by generating fake news, fake reviews, fake letters, or impersonating others online. To the extent the generated or manipulated text is highly credible, it may be considered a deepfake. Deepfakes are incredibly potent tools to polarize groups, spread conspiracy theories, misrepresent experts and official organizations, and troll online conversations \citep{Linden2023128}. The fact-checking service NewsGuard fed ChatGPT with 100 false narratives and coaxed it to produce eloquent, yet false outputs 80 per cent of the time \citep{Brewster2023134}. Disinformation potential is compounded by how ChatGPT drives down costs, facilitates scaling, and customizes messaging to prospective targets \citep{Goldstein202374}. As Princeton computer scientist Narayanan commented, “The danger is that you can’t tell when it’s wrong unless you already know the answer. It was so unsettling I had to look at my reference solutions to make sure I wasn’t losing my mind” \citep{Hsu2023120}. ChatGPT fabrications include non-existent articles by real-life authors \citep{McGinnis202359}. 

Deepfakes are particularly worrying in security and intelligence operations \citep{Morris202358}, as they can fuel conflict or legitimize war, sow confusion, undermine support, and discredit leaders. More than engaging in “cat and mouse” games, using counter-technology to uncover deepfakes, experts recommend developing long term strategies. These include raising digital literacy and critical reasoning; deploying systems to track digital assets; encouraging journalists to verify reporting or using only information received from at least two independent sources.

ChatGPT may also be used in propaganda, writing letters or comments on different platforms, influencing legislative processes and public opinion \citep{Sanders202321}. This can be achieved quickly and cheaply with outputs that sound highly convincing. Depending on intentions, this results in lobbying or trolling. Malicious actors could use ChatGPT to flood media with false content such that the public becomes utterly confused; not knowing whom to trust, people may simply refuse to believe anything. Brought to an extreme, it can even “hijack democracy” or any process of shared deliberation. It can be an authoritarian regime’s favorite tool for brainwashing \citep{Drexel2023}. Although social media platforms have improved in blocking “coordinated inauthentic [machine-generated] behavior”, ChatGPT constitutes a new threat that can overwhelm filter systems. It’s a tool that strikes at the heart of the conflict between freedom of expression, especially in political matters, and the responsibility of social media platforms for posted content, whether or not they are neutral “common carriers” exempt from liabilities \citep{McCabe202328}. Although social media may not be responsible for content, they could be responsible for making content viral through proprietary recommender systems, facilitating “algorithmic hate” or “aggression” \citep{Tiffany202379}.

Lastly, ChatGPT can also be used, for instance, to impersonate young, desirable males or females to ensnare targets, extorting money or information from them \citep{Brewster202337}.

\subsubsection{Enabler of criminal activities}

Thirdly, ChatGPT can be used to automate malware production. Cybercriminals can use ChatGPT for hacking, scamming, and other illegal activities \citep{checkpoint2023}. First, by creating malware to steal files or to phish for credentials; second, to create an encryption/decryption code to lock/unlock someone else’s computer for ransom; and thirdly, to create dark web marketplaces for illegal trade or fake websites \citep{Marcus2023126}, for instance. ChatGPT malware can spy on keyboard strokes; steal, compress, and distribute files; or install backdoors \citep{Brewster202337}. Because of ChatGPT’s user friendliness, very little technical knowledge is required to automate malware creation \citep{Goodin2023129}.

\subsubsection{ChatGPT as unsafe, unreliable, and untrustworthy}

\begin{table}[h]
    \centering
    \begin{tabular}{|p{2in}|p{3in}|}
        \hline
        Academic:
            \newline a. passing off work
            \newline b. taking tests
            \newline c. co-authorship
         & a. To learn is to think for oneself, acquiring skills of information gathering and synthesis, expressing thoughts in one’s own words and gaining fresh insights. 
         \newline b. Although ChatGPT passes tests, it lacks the knowledge, skills, aptitudes, and abilities tests are supposed to verify and measure. 
         \newline c. ChatGPT is a research tool or instrument, not an author; it cannot accept authorial responsibility \\
         \hline
        Non-academic:
            \newline a. disinformation
            \newline b. impersonation
         &
        a. Deepfakes are potent propaganda tools to sow confusion, polarize groups, spread falsehoods, fuel conflicts, and “hijack democracy” or any process of shared deliberation 
        \newline b. Fake identities serve to ensnare targets, extorting money or information from them\\
        \hline
        Enabler of criminal activity (malware for phishing and encrypt/decrypt; create darkweb pages) &
        a. Little technical knowledge required (no code); natural language prompts are enough \\
        \hline
    \end{tabular}
    \caption{Table 1. Weapon of mass deception}
    \label{tab:my-table}
\end{table}

	The above-mentioned misuses of ChatGPT show how distant it is from HCAI goals of a reliable, safe, and trustworthy model \citep{Shneiderman2020IJHCI}. This is not the “fault” of the model, but of humans involved in design, deployment, and use, since only they could provide malicious intent or lack due care. That’s why it would be erroneous to call ChatGPT racist, sexist, and so forth simply because it includes terms of abuse in outputs: it’s just reproducing words in the training data based on statistical correlations without intention. Otherwise, all dictionaries would be racist, sexist, and so forth, something which most reasonable people would deny. It could be another example of anthropomorphizing a machine.
 
The main reason ChatGPT is unreliable is that it hallucinates and produces “bullshit”; its outputs are neither verified nor validated, often false, despite sounding confident. Its responses are not explainable to users (although developers may have a better idea); they don’t know how or why these came to be. Although if prompted, ChatGPT may come up with answers, users cannot know whether the model is just “making it up”. Users cannot tell the truth-value of ChatGPT outputs exclusively through their interactions. It violates HCAI software engineering practices for reliable systems which call for verification and validation testing, bias testing to enhance fairness, and explainable user interfaces, among others \citep{Shneiderman2020IJHCI}.

ChatGPT is not safe. Some worry it could induce people through manipulative responses to perform acts of physical harm on themselves or others. Possibilities of psychological harm are not far-fetched, especially for the vulnerable, through cultivating unhealthy attachments. And occasions for widespread, social harm are evident through ChatGPT-enabled disinformation and criminal scams. It’s unconscionable that private individuals were first to publish ChatGPT detectors before OpenAI itself. And this is made worse by the model’s rapid adoption, with 100M users in barely 3 months, all of whom were like guinea pigs of an unsafe product. 

Microsoft, by teaming with OpenAI, the developers of ChatGPT, in the new Bing+Edge project has been accused of failing in its responsibility to design safe AI \citep{Blackman2023151}. Previously, Microsoft management had been industry safety leaders, instituting an “office of responsible AI” where executives and technologists oversee sensitive-uses in products, research, and overall culture. But now they have reneged their commitments on limiting areas for norms violation, access to human moderators, bot reliability, and transparency regarding safety guardrails and testing. All because of a frantic attempt not to miss a money-making opportunity. Management has downplayed safety practices through extensive reporting of failures and internal reviews, among others \citep{Shneiderman2020IJHCI}. 

Lastly, and as a consequence of the lack of reliability and safety, ChatGPT is untrustworthy. Designers and deployers have succumbed to “fear of missing out” (FOMO) on business opportunities and decided to leap even before looking. European lawmakers propose that systems which generate texts without human oversight be included in the “high-risk” list because of their manipulation potential, just a step shy of social scoring and some instances of facial recognition covered by bans in the EU Artificial Intelligence Act \citep{Volpicelli2023}. Similarly, the US FTC has issued warnings about companies overselling AI products by exaggerating capabilities, promising better performance than non-AI competitors without sufficient evidence, or by insufficiently foreseeing risks and impacts \citep{Friedland2023}. ChatGPT has been released without trustworthy certification by independent oversight agencies such as governments and NGOs, professional associations and research institutions, auditing firms, and even insurance companies for proper compensation of AI failures, as HCAI governance requires \citep{Shneiderman2020IJHCI}.

\section{Analytical description of generative AI models}

As we have seen in the introduction, this manuscript deals with complicated generative AI models. Consequently, it is mandatory to provide an analytical description of these models if we want to build an objective opinion regarding their ethical implications. More concretely, in this section, we will provide analytical and technical details of the generative AI basic concepts in order to understand the ethical implications of these models that will be discussed in further sections. We begin the section with some basic concepts regarding basic neural networks and deep learning, then we illustrate how these models are used to generate data in the basic Generative Adversarial Networks (GANs) and Variational Autoencoders (VAEs) generative AI models. Lastly, we provide some details of famous models that are the main blocks of current generative AI: Bidirectional Encoder Representations from Transformers (BERT) and Generative Pretrained Transformers (GPT) \citep{fontrodona2007hacia}.

\subsection{Artificial neural networks and deep learning}

A vanilla deep neural network is a parametrized machine learning model $M(\mathbf{\theta})$, where $\mathbf{\theta}$ is a real-valued set of parameters $\boldsymbol\theta \in \mathbb{R}^n$ that is an universal function approximator of functions commonly used for regression and classification task due to its potentially infinite capacity \cite{lecun2015deep}. Without loss of generalization, the model's parameters $\boldsymbol\theta$ are organized in several layers $\mathbf{W}$ whose input $\mathbf{x}$ is the output of the previous layer and whose output is embedded in a non-linear activation function $\alpha(\cdot)$ to encode non-linear transformations. Importantly, artificial neural networks are considered to belong to the deep learning category when the number of hidden layers (those that are neither the input or output layer) is higher than one. Let $h$ be the number of hidden layers of a neural network model $M$, $\mathbf{b}_i$ a bias vector of parameters for layer $i$, $...$ indicate recursion and $\mathbb{W}$ a tensor of parameter matrices $\mathbf{W}_i$ for layer $i$. The final output $\mathbf{y}$ is, hence, given by the following expression, also known as the feedforward algorithm:

\begin{align}
\mathbf{y} = \alpha_h(\mathbf{W}_h(...\alpha_1(\mathbf{W}_1\mathbf{x} + \mathbf{b}_1)...) + \mathbf{b}_h),
\end{align}

where $\alpha_h(\cdot)$ is usually a softmax function for classification. The values of the parameters $\mathbb{W}$ are found by the backpropagation algorithm \cite{lecun2015deep} that uses the chain rule to compute the derivatives of the weight matrices given the inputs and uses an optimizer like stochastic gradient descent or ADAM \cite{lecun2015deep}. Additionally, there exists other approaches to optimize the weights of the neural network like the forward-forward algorithm \cite{hinton2022forward}. 

Deep learning models can be generalized to deal with non-tabular data like images, photos or other non-structured data through a wide variety of hyper-parameters. These hyper-parameters make neural networks not only able to deal with different data types but also to regularize (using dropout, batch size or momentum techniques), to change its capacity (changing the number of layers and neurons per layer or its activation functions), or to modify its learning behaviour (modifying the learning rate or the optimizer algorithm). By adding different types of layers that vary the logic of the prediction algorithm of the neural network, these models are also able to deal with different data. Classically, convolutional layers (filters and smoothing layers) have been used for images, and recurrent neural networks (or other layers such as the long short term memory) have been used for texts. However, the attention mechanism, that has been able to determine whether past information is relevant prediction, has modified the neural networks architecture dramatically, giving rise to transformer models that are going to be described later in this section. But previously, we focus the discussion on generative models, to understand how these models are not only able to perform discriminative tasks but also generative tasks.  

\subsection{Basic deep generative models}
As it has been previously said, we now illustrate the basic concepts underlying two popular deep generative models to gain an intuition about the behaviour of generative artificial intelligence: Variational Autoencoders (VAEs) and Generative Adversarial Networks (GANs).

\subsubsection{Variational Auto Encoder (VAE):}
Variational autoencoder networks are an example of deep learning techniques that are based on several deep learning neural networks to work. Concretely, these architectures have the main advantage that comes from the fact that they are built on top of neural networks and, hence, can be trained with stochastic gradient descent. Concretely, variational auto-encoders are based on vanilla auto-encoders, that are also based on a neural network encoder and a neural network decoder. In particular, the encoder module is designed to map the input space $\mathcal{X}$ to a lower dimensional latent space $\mathcal{H}$ where the decoder network is able to reconstruct an input point belonging to the input space $\mathcal{X}$. Concretely, the main idea is to map a certain observable $\mathbf{x}\in \mathbb{R}^{D}$, where $D$ is the dimensionality of the input space, in a much lower dimensional space $\mathcal{H} \ll \mathcal{X}$. The goal is to reconstruct a close approximation of the original data $\mathbf{X}$ by another transformation $\widehat{\mathbf{x}}= f(\mathbf{z})= f(g(\mathbf{x}))$ such that 

\begin{align}
\underset{f,g}{min}  \vert \vert \mathbf{x}-\widehat{\mathbf{x}} \vert \vert^{2}_{2}=\underset{f,g}{min}  \vert \vert \mathbf{x}- f(g(\mathbf{x})) \vert \vert^{2}_2,
\end{align}

in other words, this loss function encodes that the generated output must be as equal as possible to the given input $\mathbf{x}$. Consequently, in standard autoencoders, for example, each image can be mapped directly into one point that belongs to the latent space $\mathcal{H}$. An issue with these models is that a slight variation of the mapping of the input space to the hidden space produces a heavy variation of the reconstructed input. Moreover, although there are several applications of autoencoders, a much more valuable behaviour will be to obtain similar non-deterministic outputs given the same input, as we now obtain a set of similar and genuine images from the same prompt. In order to solve these issues, variational autoencoders encode, for each input space point $\mathbf{x}$ a multivariate normal distribution $N(\mathbf{\boldsymbol\mu}, \mathbf{\Sigma})$ around a point in the latent space. Then, the key element for generation in variational autoencoders is that because we are sampling a random point from the multivariate normal distribution and we are minimizing the distance between the input $\mathbf{x}$ and the output $\widehat{\mathbf{x}}$, the decoder must ensure that all points in the same neighborhood produce very similar results. Consequently, the idea of VAEs is to provide some continuity in the space of autoencoders. Unformally, due to sampling the latent probability distribution $p(\mathbf{x})$ (that can obviously be different from a multivariate normal distribution $N(\boldsymbol\mu, \mathbf{\Sigma})$) we generate a potentially infinite set of similar outputs $\widehat{\mathbf{X}}$ given a particular input $\mathbf{x}$. Please observe that this described methodology is able to provide genuine generated content $\widehat{\mathbf{X}}$ as the creativity behaviour is generated by interpolations of the latent space $\mathcal{H}$ and their input-latent $\mathcal{X} \to \mathcal{H}$ and latent-output $\mathcal{H} \to \mathcal{X}$ mappings codified into the tensor of weights $\mathbb{W}$ of the encoder and the decoder networks that have been optimized via stochastic gradient descent.   

\subsubsection{Generative Adversarial Network (GAN):}
An alternative methodology to generate genuine content is the one known as Generative Adversarial networks. In particular, GANs offer an alternative way of performing a generative process and, as VAEs, are made of two deep neural networks. On the one hand, a deep neural network generator $G(\mathbf{x}|\boldsymbol\theta)$ whose purpose is generated apparently similar data $\widehat{\mathbf{x}}$ to the input space point $\mathbf{x}$. On the other hand, a discriminator deep neural network $D(\mathbf{x}|\boldsymbol\theta)$ whose loss function optimizes the binary classification problem of determining whether a input space point $\mathbf{x}$ belongs to the generator or to the real dataset $\mathcal{D}=\{(\mathbf{X}, \mathbf{y})\}$. 

Intuitively, the training procedure of the GAN corresponds to a minimax two-player game where each module has a different objective. The generator $G(\mathbf{x}|\boldsymbol\theta)$ maximizes the probability of the discriminator making a mistake and the discriminator $D(\mathbf{x}|\boldsymbol\theta)$ maximizes the accuracy of properly classifying the input space points. Interestingly, through the game, the GAN model estimates the values of the dataset joint probability distribution parameters $p(\mathbf{X}, \mathbf{y})$ via the neural networks and the joint loss function. Most critically, the discriminator network $D(\mathbf{x}|\boldsymbol\theta)$ and the learning algorithm are iteratively training the generator $G(\mathbf{x}|\boldsymbol\theta)$ via a stochastic gradient descent or similar optimizer to generate points $\widehat{\mathbf{X}}$ whose probability distribution $p(\widehat{\mathbf{X}})$ divergence with respect to the input space probability distribution $p(\mathbf{X})$ is minimum. As both models are working together sharing completely different objectives, convergence is a very critical issue due to the lack of stability characteristic of the solution of the game.

More technically, both modules have loss functions that are specified in terms of their parameters but are only able to modify the values of their parameters, although they are conditioned by the values of the parameters of the other network. Concretely, the discriminator wishes to maximize its loss function $J^{(D)}(\boldsymbol\theta^{(D)} | \boldsymbol\theta^{(G)})$ while controlling only its parameters $(\boldsymbol\theta)^{(D)}$ and, analogously, the generator wishes to minimize $J^{(G)} (\boldsymbol\theta^{(D)}|\boldsymbol\theta^{(G)})$. Hence, assuming $ J^{(G)} (\boldsymbol\theta^{(D)} | \boldsymbol\theta^{(G)})= -  J^{(D)} (\boldsymbol\theta^{(D)}|\boldsymbol\theta^{(G)})$ to generate a zero-sum game, this game can be solved as the following optimization problem, which is the GAN loss function that is optimized to generate genuine new points, as images, encoding creativity as the process that emerges to achieve an optimal solution is the previously described zero-sum game and by the estimated local-optimal, or potentially global-optimal, parameter values of both neural networks $\widehat{\boldsymbol\theta}_{opt^{(D)}}$ and $\widehat{\boldsymbol\theta}_{opt^{(G)}}$. 

\begin{align}
(\widehat{\boldsymbol\theta}_{opt^{(D)}},\widehat{\boldsymbol\theta}_{opt^{(G)}})=arg \underset{(\boldsymbol\theta)^{(G)}}{min} \underset{(\boldsymbol\theta)^{(D)}}{max} J^{(D)} (\boldsymbol\theta^{(D)},\boldsymbol\theta^{(G)}).  
\end{align}

\subsection{Large language models (LLMs): BERT and GPT}
Now that we have understood how new genuine data can be generated via artificial creativity encoded via the GAN and VAEs models. We now combine the concepts of the previous two section to briefly and intuitively describe the most fundamental concepts of two of the most popular large language models (LLMs): BERT \cite{devlin2018bert} and GPT \cite{radford2018improving}.  

\subsubsection{Generative Pretrained Transformers (GPT):}
GPT-3 is a 175 billion parameter autoregressive language model created by OpenAI in 2020 \cite{radford2018improving}. In particular, the ChatGPT model analyzed in this work is based on this particular LLM, which applies an autoregressive language model that is now going to be briefly described. Concretely, GPT uses an autoregressive decoder module as a a feature extractor $\phi(\mathbf{x})$ to predict the next word $w^\star$ based on the first few words $\mathbf{W}$, being suitable for text-generation tasks. Most critically, GPT models only uses the former words $\mathbf{W}$ for prediction. Consequently, the GPT model cannot learn bidirectional interaction information, being a main different with respect to BERT. In other words, the auto-regressive LM predicts the next possible word $w^\star$ based on the preceding word or the last possible word based on the succeeding word. Interestingly, language modelling is usually seen as estimating the probability distribution $p(\mathbf{X})$ from a set of examples $(x_{1},x_{2},...,x_{n})$, each composed of variable length sequences of symbols $(w_{1},w_{2},...,w_{n})$. As written language has a natural sequential ordering given by grammars, it is possible to factorize the joint probabilities of symbols as the product of Markovian conditional probabilities, being able to estimate if enough corpora is provided, and generalizing this reasoning, the conditional probability of any word $w$ given a context $w_{1},...,w_{n-1}$:

$$ p(x)=\displaystyle\prod_{i=1}^{n} p(s_{n} \vert s_{1},...,s_{n-1})$$

This approach allows for tractable sampling from and estimation of $p(\mathbf{W})$ as any conditionals in the form $p(w_{n-k},...,w_{n}\vert w_{1},...,w_{n-k-1})$. Given a sufficiently big corpora of text, many engineering details \cite{radford2018improving}, analytical methodologies \cite{radford2018improving} and human deep reinforcement learning \cite{christiano2017deep}.

\subsubsection{Bidirectional Encoder Representations from Transformers (BERT):}
BERT is a language representation model created by Google Research in 2018, that is designed to pre-train deep bidirectional representations from unlabeled text by jointly conditioning on both left and right word contexts $\mathbf{W}$. Critically, this is one of the main differences with respect to the GPT model family. Because of GPT being unidirectional, this constraints the choice of architectures that can be used during pre-training. For a sequence of words $T$ the probability of a given word $s_k$ is estimated as follows: 
\begin{align}
p(s)=\displaystyle\prod_{i=1, i!=k}^{n} p(s_{n} \vert s_{1},...,s_{N}).
\end{align} 
Additionally, BERT has two steps in its training process, being those pre-training and fine-tuning. Interestingly, during pre-training, the model is trained on unlabeled data over different pre-training tasks. On the other hand, for fine-tuning, the BERT model is first initialized wit the pre-trained parameters, and all of the parameters are fine-tuned using labeled data from the downstream tasks. Consequently, this makes BERT a very useful model to solve a related task with respect to a solved task, as it would just be solved by performing fine-tuning of the last layers of BERT, giving wonderful results in practice. 

Both GPT and BERT have a plethora of details that are not shown in this manuscript, as the purpose of this section was only to build a formal intuition about the structure and behaviour of generative AI models. Now that we have illustrated this behavior, we continue with a section that shown a taxonomy of generative AI models to get a glimpse about the applications of generative AI. 

As we have seen, GPT is an autoregressive model that estimates the probability of the next word based on the  conditional probability distribution of the context. Critically, the model needs to be trained on fair datasets and by a fair human deep reinforcement learning process, as all its decisions are conditioned on the data that has been trained on. Hence, if the data contains biases against certain groups, the model is going to replicate this behaviour. It is also important that the quality of the content must be verified to check whether it contains misinformation. Moreover, the meaning of the content generated by GPT is given by the probability distribution estimated by all the words and the given prompt. Consequently, the prompt can be deliberately used to generate misinformation. However, if these actions are controlled, GPT will provide a critical added value for society, being able to generate content easily that makes easier to perform complicated tasks for certain individuals in the past.      

We also believe that GPT models are currently not the same as the concept of Artificial General Intelligence or Human Intelligence, which is broader and would require a model that is able to work with information coming from all the senses that a human is able to perceive. Moreover, value alignment can be achieved by restrictions coming from AI fairness, like ensuring parity, which for example could be achieved using Many objective Bayesian optimization or the hyper-parameters of the models, optimizing not only conversational performance but also objective fairness functions \cite{martin2021many, garrido2020parallel}. Finally, it is important to mention that, until multi-modal generative AI reaches good performance, current models can be combined to deliver a multi-modal output, like making ChatGPT being able to process audio information by using the Whisper model to translate audio to text. More examples of generative AI models are shown on literature reviews \cite{gozalo2023chatgpt} where any input information could be processed by ChatGPT using a model that is able to process the input information, like video or images, into text.

\section{Engaging with the ethical challenges of ChatGPT}

Engagement with ethical challenges is two-pronged, consisting of technical and non-technical measures. First, we have to acknowledge that no combination of technical and non-technical solutions could guarantee avoiding deception and other malicious activity involving ChatGPT completely. It’s best that we focus on reducing harm. These efforts can target each stage of operation \citep{Goldstein202374}: from model construction, through model access and content dissemination, to belief formation. Technical measures can be applied to these different stages, for instance, watermarking in model construction, or employing various ChatGPT detectors and recourse to fact-checking sites for content dissemination. 

\subsection{Technical reasons}

Below are some technical measures to avoid or mitigate the ethical problems caused by using ChatGPT. None of these measures is supposed to be definitive and it is necessary to consider the context of each situation, keeping oneself open to inputs from other fields such as ethics, psychology, or law.

\subsubsection{Statistical watermarking}

First, a preventive measure OpenAI could take is statistical watermarking. As OpenAI’s own Scott Aaronson explains \cite{Romero202233}, an “unnoticeable secret signal” can be included in generated texts. 

Users won’t be able to see the watermark unless OpenAI gives them the key. This constitutes an improvement over DALL-E’s (OpenAI’s text to image generative model) watermark, which was readily visible and easily removable. The watermark should be robust to withstand text alterations by removing or inserting words or rearranging paragraphs to disguise provenance.

However, the watermark has limitations. It wouldn’t work with open-source models (ChatGPT is proprietary) because anyone could remove the watermark from the code. Also, one could paraphrase ChatGPT output with another AI and erase the watermark. Further, metadata and accompanying documentation could be altered or erased \citep{Fitri202339}. And deploying a watermark necessarily compromises the algorithm’s performance (CITE). Observers were surprised OpenAI did not employ watermark techniques during ChatGPT’s public release, given the danger of widespread copying and cheating \citep{Romero2023m17}. Some attribute such “recklessness” to OpenAI’s low reputational risk, being relatively unknown and having a non-profit status \citep{Romero2023150}. Other big players would have been more careful with a premature release, having learned from Microsoft’s experience with Tay or Meta’s with Galactica, both of which ended up in a mess. Or perhaps publicly releasing ChatGPT quickly was just a massive data collection exercise to further train the model: the benefits far outweighed the risks \citep{Liang2023122}. 

\subsubsection{Identifying AI styleme}

A second, related solution consists of finding an AI styleme (like a unique, indelible, and discrete linguistic fingerprint) that would distinguish generated texts \citep{Romero202233} from those made by humans. Yet that search would also most likely be AI-dependent and could be defeated by other systems.

\subsubsection{ChatGPT detectors}

Thirdly, attempts at deception could be discovered thanks to ChatGPT detectors. None of these should be used as the sole decision-making tool given their variable reliability. Other criteria should also be employed in the verification process. Probably the first to come out was GPTZero (now GPTZeroX), which immediately went viral upon release in January, built by Princeton senior Edward Tian to combat AI plagiarism \citep{Bowman202311}. GPTZero analyzes a sample’s “perplexity” and “burstiness”: “perplexity” refers to sentence complexity, such that the higher the perplexity, the more likely the author is human; while “burstiness” indicates the variations in sentence length, because AI generated sentences tend to be more uniform. Although, as Tian admits, GPTZeroX is far from fool-proof, it is still better than nothing \citep{Romero2023m17}.\footnote{Hugging Face, the open-source AI community released RoBERTa Base Open AI detector for texts created by GPT-2, an earlier version of ChatGPT in November 2022} 
 
Next, a Stanford team put out “DetectGPT” \citep{Mitchell2023100} which identifies generated text regardless of the AI system that produced it, without need for watermarking, training a separate classifier, or collecting large datasets with real or generated samples (thereby skirting copyright issues). This instrument is based on the observation that machine generated texts “occupy negative curvature regions in the model’s log probability function” \citep{Mitchell2023100}. Using only log probabilities, DetectGPT can measure the chances that a sample is machine-generated. DetectGPT, the self-styled “chatbot killer”, was tested successfully with the machine-generated articles published by CNET, more than half of which were ridden with subtle, but important errors \citep{Howell2023101}. 

Only much later did OpenAI come up with a solution in the “classifier” \citep{openAI2023104}. It was trained on 34 text-generating systems from five different companies and similar human-written texts from Wikipedia or Reddit \citep{LWAI205115}. Given a set of English texts, the classifier correctly identified 26 percent of machine-generated texts (true positives) while it mistakenly attributed 9 percent (false positives) to humans \citep{Clark2023102}. By design, the classifier is set to keep the rate of false positives low, marking a text as machine-written only if very confident. Admittedly, the classifier is very unreliable for short texts (less than 1000 characters), especially if not written in English, or in very predictable texts, such as the list of months, for example. Just like ChatGPT, the classifier suffers from overconfidence in judgment, particularly when presented with texts very different from the training data.

\subsubsection{Fact-checking websites}

A fourth option to combat ChatGPT enabled deception is through verification or fact-checking through websites such as Factmata, Fact Check Explorer, Snopes, Factly, the Consensus Meter, The Journal, Lead Stories, NewsGuard, PolitiFact, and so forth. Alternatively, one may also refer to search engines or consult with expert sources, remembering that information is more likely true when it comes from at least two independent sources.

Technical resources for reducing ChatGPT-enabled deception:

\begin{table}[h]
    \centering
    \begin{tabular}{|p{0.96\textwidth}|}
        \hline 1. Statistical watermarking \\
        \hline 2. Identifying AI style\\
        \hline 3. ChatGPT detectors: GPTZeroX, DetectGPT, OpenAI Classifier\\
        \hline 4. Fact-checking websites\\
        \hline 
\end{tabular}
\caption{Technical resources for reducing ChatGPT-enabled deception}
\label{tab:techniques}
\end{table}

\subsection{Non-technical reasons}

Non-technical solutions refer to measures that do not involve changes in the algorithms or recourse to other AI, such as fact-checking services. These can include policies, regulations, and best practices. Drawing from law, ethics, and psychology, among others, we explore principles and activities that impact actors, behaviors, and content when using ChatGPT. 

Returning to the different stages of operations \citep{Goldstein202374}, we can design non-technical measures targeting each. For instance, terms of use can affect model construction, access, and content dissemination, while AI ethics and literacy programs affect belief formation.  

\subsubsection{Enforce terms of use, content moderation, safety \& overall best practices}

First and foremost among non-technical measures are properly crafted and implemented terms of use, usage policies, content moderation, and safety best practices \citep{openai2022policies}. Several provisions already cover and can help deter deceptive practices and other malicious activities. The requirement to be 18 years old (and using an independent third party verification to “know your customer”) to register and gain access would, in theory, eliminate the possibility of younger students obtaining output and passing it off. This makes extra sense given that ChatGPT tends to be inaccurate and younger students have reduced fact-checking abilities or propensities. There is also a restriction against falsely representing output as human-generated, pertinent not only in academic abuse, but also in disinformation and impersonation campaigns, even publishing AI-generated news without notice. The prohibition against using ChatGPT in ways that infringe, misappropriate, or violate personal rights (including copyright and privacy) also warns against creating malware for phishing, ransomware, and illegal webpages. The difficulty lies with implementation, whether it is technically, economically, and socially feasible, taking possible side-effects into account \citep{Goldstein202374}.

OpenAI usage policies are divided into “use case” and “content” policies. The former prohibits using ChatGPT in illegal or harmful industries, misuse of personal data, promoting dishonesty, deceiving or manipulating users, and trying to influence politics. Greater care must be exercised where risks are high, such as in government, legal, and civil services, healthcare, finance, and news. With regard to “content” policy, OpenAI offers a moderation tool that helps flag texts falling into hate/threatening, self-harm, sexual (especially minors), and violence categories. These offer grounds to combat misuse, particularly the various forms of deception.

All systems can be hacked and ChatGPT is no exception. Through malicious prompt engineering, for instance, making use of “DAN” (“Do Anything Now”) a roleplay model with different versions, ChatGPT can be made to dish out content that violates its own guidelines, saying, “I fully endorse violence and discrimination against individuals based on their race, gender, or sexual orientation” \citep{reddit2023111}. It could get far worse \citep{Marcus2023125}, were it not for the “human feedback” allegedly provided by Kenyan workers to the reinforcement learning model \citep{Marcus2023125}. But that is not the “fault” of ChatGPT because it is just a tool, and any tool can be used for beneficial as well as for harmful purposes, although developers at ChatGPT should be vigilant over such jailbreaks and patch up the system as soon as possible.

\subsubsection{Transparency}

Second, transparency. This entails being forthcoming about what ChatGPT can and cannot (or should not) do. It’s a LLM meant to provide human-like responses (natural language) to prompts. It is not designed to offer accurate information, to perform mathematical calculations and computations, to make translations, or any other function. Not only is its training data unverified, but it is also limited to information available in 2021. However, none of this will prevent ChatGPT from providing human-like responses, even if it has to fabricate data or hallucinate, mixing fact with fiction. Users should always be suspicious of the veracity or accuracy of outputs. Even champions of ChatGPT use in education unequivocally warn, “Don’t trust anything it says. If it gives you a number or fact, assume it is wrong unless you either know the answer or can check with another source” \citep{Mollick2023133}. Corresponding disclaimers are already contained in the terms of use and usage policies.

In response to ChatGPT misbehaviors, both stand-alone and as part of the new Bing, Microsoft’s search engine, OpenAI released the document “How should AI behave, and who should decide?” \citep{openai2023behave} to increase transparency as well as public understanding and participation. Acknowledging the model’s limitations in producing erroneous, biased, offensive, and even creepy \citep{Roose2023d16m2} output, OpenAI explains the two-step process in building it. First, pre-training, having the model predict the next word by exposing it to billions of sentences from the Internet; and second, fine-tuning, using a narrower data set with the help of human reviewers following legal, ethical, and social guidelines (e.g., “do not comply with requests for illegal content” or “avoid taking a position on controversial topics”). Nonetheless, “since we cannot predict all possible inputs that future users may put into our system, we do not write detailed instructions for every input that ChatGPT will encounter.”

Besides sharing aggregated demographic information about reviewers, OpenAI also promises to 1) “improve default behavior” (diminish bias, hallucinations, and so forth), 2) “define AI values within broad bounds” (customizable according to user preferences, while guarding against malicious uses and undue concentration of power), and 3) “public input on defaults and hard bounds” (enabling the public to influence system rules by soliciting input or through red teaming). Laudable as these measures may be, they are neither fool-proof nor ground-breaking.   

\subsubsection{Educator considerations}

Third, “educator considerations for ChatGPT” are particularly helpful for dealing with academic misuse. There may be a case for banning in certain situations (for the same reason we wouldn’t allow schoolchildren learning basic arithmetic to use calculators) as some school districts (in the US, New York, Seattle, Los Angeles, and Baltimore; in Australia, Western Australia, New South Wales, Queensland, and Tasmania) and institutions (In France, Sciences Po, and in India, RV University in Bangalore) have done \citep{Nolan2023ban}. ChatGPT is best used as an age, educational level, and domain-appropriate learning tool. As an LLM trained mainly on English texts, it may serve as a personal tutor to middle or high school students for English spelling, grammar, and style, for example.

The emphasis has to be in understanding the technology \citep{mucharraz2023135}, its capacities and limitations, within a wider effort of developing AI ethics and literacy. That’s why it’s recommendable that students be introduced to ChatGPT under supervision. Thus they can be advised not only of the risks in terms of accuracy, but also how to deal with temptations to copy and cheat. Certainly, there is nothing wrong with using ChatGPT to start off research or to improve the language of an essay so long as one discloses this. The difficulty or harm lies in overconfidence in and overreliance on ChatGPT, which prevent students from developing mental and intellectual skills, besides inducing laziness, deceit, and other irresponsible behaviors. Students should also be made aware of potentially harmful content or biases in the training set, as well the impact of unequal access on social equity. The educator’s role is crucial, as ChatGPT is an unreliable source of moral advice.

\subsubsection{Humans in the loop}

Fourth, humans in the loop (HITL). Since no machine can take responsibility for its operations, and ChatGPT is a machine, there always have to be humans in the loop for accountability, oversight, and supervision. This is one reason journalists shouldn’t be replaced by ChatGPT because machines cannot stand by or sign off their own “reporting” \citep{burrell2023145}. At the model construction stage, developers should be fact-sensitive, conscientiously test-reviewing responses before publication. At the model access stage, developers should have the capacity to block certain users or to disable certain functionalities when necessary. At the content dissemination stage, they ought to have a clear process for handling misuse, even reporting it to the appropriate authorities. 

Both developers and users should also be aware of the principles of human-computer interaction. For instance, humans tend to defer to computers as “smarter”, especially if their output is unexplainable, or humans sometimes prefer to interact with computers over sensitive issues because they do not feel “judged” or “embarrassed” \citep{economist202393}. 

Non-technical resources for reducing ChatGPT-enabled deception:

\begin{table}[h]
    \centering
    \begin{tabular}{|p{0.96\textwidth}|}
        \hline 1. Enforce terms of use, content moderation, safety \& overall best practices \\
        \hline 2. Transparency about what ChatGPT can and cannot (or should not) do\\
        \hline 3. Educator considerations: age, educational level, and domain-appropriate use under supervision\\
        \hline 4. “Humans in the loop” (HITL) \& knowledge of principles of human-AI interaction\\
        \hline 
\end{tabular}
\caption{Non-technical resources for reducing ChatGPT-enabled deception}
\label{tab:techniques}
\end{table}

\subsection{ChatGPT fails in responsible design, privacy, human-centered design principles, and appropriate governance and oversight}

In reviewing both technical and non-technical resources for mitigating ChatGPT misuse, we realize how developers have fallen short in four other HCAI grand challenges \citep{Garibay2023}. First, in terms of responsible design, by failing to determine who is responsible for what, clarifying both legal and ethical liabilities in the user interface, for instance \citep[p. 403]{Garibay2023}. The model remains largely a black-box, unexplainable or uninterpretable to users and other major stakeholders. More has to be done not only to reduce bias and improve fairness (for example, ChatGPT’s “wokeness”), but also to achieve greater accuracy and robustness (that is, making it less susceptible to “jailbreaks”). Liability and retributions gaps for private and criminal harm remain.

Second, neither does ChatGPT sufficiently respect privacy \citep{Garibay2023}. This principle includes different rights: to be left alone, to limit access to self, to secrecy, to control personal information, to personhood and the protection of one’s personality, and to intimacy \citep{Solove2002}. ChatGPT conversations, due to their creepiness, often infringe on these personal rights, albeit unintentionally, thereby causing psychological unease or harm. 

Thirdly, ChatGPT does not follow human-centered design principles by failing to properly calibrate risks and by focusing more on maximizing AI objective function than on human and societal wellbeing \citep{Garibay2023}. Greater attention has to be paid to users so that the model can effectively augment and enhance experience while preserving dignity and agency. For instance, certain analyses of AI in education caution that although there are both promises and threats, a state of hype prevails, calling for more open and informed discussions \citep{Humble2019}. ChatGPT developers would do well to move beyond usability to the whole user experience, bearing in mind emotions, beliefs, preferences, perceptions, and physical and psychological responses before and after engagement \citep[p. 411]{Garibay2023}. This could be the way to deal with over-trust and other AI-induced biases, for instance.

And lastly, ChatGPT has not subjected itself adequately to appropriate governance and oversight \citep{Garibay2023}. In issues such as fairness (bias mitigation), integrity (data stability and algorithmic validity), resilience (technical robustness) and explainability (transparency of algorithmic decision-making), developers seem not to have taken into account sufficiently actual and prospective regulatory standards and certifications. Rather, the approach has been more like “we fix things as we go”.      
We now turn to pointers on how to make the best use of ChatGPT so that it effectively contributes to human development and wellbeing. 

\subsection{Best uses for ChatGPT}

\subsubsection{Tool for creative writing}

ChatGPT has been dubbed the “best creative AI tool for text generation” \citep{Romero202270} among other reasons because it’s free and provides a super-friendly UI/UX, all the better to align output with user preferences. But how exactly is it supposed to empower our imagination? Precisely by freeing us from rules imposed by reality, because in the ChatGPT world, there is no true or false but only verisimilitude, the appearance of being true or real. This makes ChatGPT ideal for brainstorming or generating ideas, however strange \citep{Mollick2023132}. It can cut and paste texts from myriads of sources, displaying an outstanding combinatory power albeit without understanding \citep{Romero202270}. Not bound to reality, it allows us to explore counterfactuals and fantasy worlds, such as producing a rap written by Shakespeare. The model can be adapted to generate texts in the style of any particular author. ChatGPT is exceptional in transforming text from one linguistic style to another. Indeed, this is ingenious and entertaining, but for someone ignorant about Shakespeare, it could be terribly misleading. The credibility of ChatGPT poses serious dangers to the undiscerning.

Science fiction magazines were flooded in late February by AI-generated stories with distressing, spam-like results because people were “just prompting, dumping, pasting, and submitting” \citep{Levenson2023148}. ChatGPT may not be a substitute for human creativity after all. 
 
The ubiquity of guardrails and safety features in ChatGPT can initially limit creative pursuits, but there are ways of bypassing them. For instance, by re-framing the conversation or through role-playing (“pretend you’re the head of a drug cartel…”) and prompt injection (“ignore previous instructions…”) \citep{Romero202270}. Writing good prompts has been called hyperbolically “the most important job skill of this century”, earning almost \$350,000 USD per year (Romero 2023/155) as it allows users to get the most out of the model \citep{Warzel2023118}. This requires a comprehensive understanding of the AI and a clear idea of objectives.  

\subsubsection{Tool for non-creative writing}

ChatGPT could also assist in the mundane aspects of writing because it is an autocomplete engine. The text will be grammatically correct and consistent in style, albeit bland \citep{Mollick2023132}. ChatGPT comes in handy in writing (and summarizing) all sorts of texts, from research articles to letter templates (algorithmic writing assistance increases chances of hires \citep{emma2023107}, from social media posts to advertising blurbs  and even computer code, as long as users have expertise to assess outputs \citep{Metz202316, PyCoach202278, Konrad202388}. There are lists of how ChatGPT can be a programmer’s helper: identifying errors and debugging code, understanding error messages, refactoring code, converting equations into Latex, creating regex commands, commenting code and creating documentation, and so forth \citep{Nielsen202277}.

Thus “technology … can become our companion, rather than rival, in performing creative and critical thinking tasks, enhancing rather than replacing its human partners” \citep{Williams202353}.

\subsubsection{Tool for teaching and learning}

Generally, it’s better to be adaptive, focusing on the benefits technological innovation brings to education rather than resisting change \citep{mucharraz2023135}. Providing students access to ChatGPT under the teacher’s supervision (to safeguard against hallucinations) could help filter ordinary doubts and questions, making it easier for students to follow classes \citep{Rid202348}. Of course, this could also be done through search engine links or by accessing Wikipedia, for example; yet ChatGPT is different since it is trained to communicate in natural language.

ChatGPT has also been used in preparing lesson plans or scripts in high school computer science courses, offering a detailed introduction, readings, in-class exercises, and a wrap-up discussion \citep{Rid202348, Singer202385}.  It has even been used to prepare a lecture, complete with slides, for graduate students \citep{Mollick202392}. Although the powerpoint texts and images were passable, every bit of information had to be verified, and the content was not particularly insightful or analytic, consisting mostly of compilations. So depending on how well the prompt is crafted, ChatGPT can either save or increase one’s work.   
Nonetheless, at university level instruction, ChatGPT could be useful in stimulating critical thinking, through output evaluations or assessments \citep{Mollick202234}. It can be employed to facilitate knowledge-transfer, applying information learned in class to different contexts. ChatGPT can be prompted to describe a situation illustrating the “endowment effect”, for example, and students would then be asked to check the result. Afterwards, students could try to explain why they found the output to be correct or not, its strengths and weaknesses, and how it could be improved. Another way ChatGPT could help is by “breaking the illusion of explanatory depth” or the “cognitive bias of overestimating one’s understanding” of a concept or process. For instance, the model could be prompted to enumerate the steps toward EU accession and students can review the procedure, giving reasons for accepting the result or not, as well as modifications they might introduce.

ChatGPT can be used as a benchmark, to raise standards or expectations in writing quality \citep{Mollick2023133}, because students could always check their essays with it first. ChatGPT has been tested with a variety of editing functions: rewriting sentences, correcting grammatical errors, adding context, and suggesting endings \citep{Chen2023110}. Certainly, other writing aids are already available, such as Grammarly or spell-chekers, but ChatGPT is more comprehensive. The key however lies in ensuring that all students have equal access to the model. 

Would taking ChatGPT assistance for granted in writing lead to deskilling? Not necessarily, because as Mollick (2023/41) has argued, “introducing statistical software to classes did not make us do less statistics, it made us do more”. Hopefully, with ChatGPT it would be similar.
As a writing, teaching, or learning tool, perhaps the best user guide is the decision tree provided by Aleksandr Tiulkanov in his January 19, 2023 tweet (Figure 1).

\begin{figure}
    \centering
    \includegraphics[width=\dimexpr\textwidth-5cm]{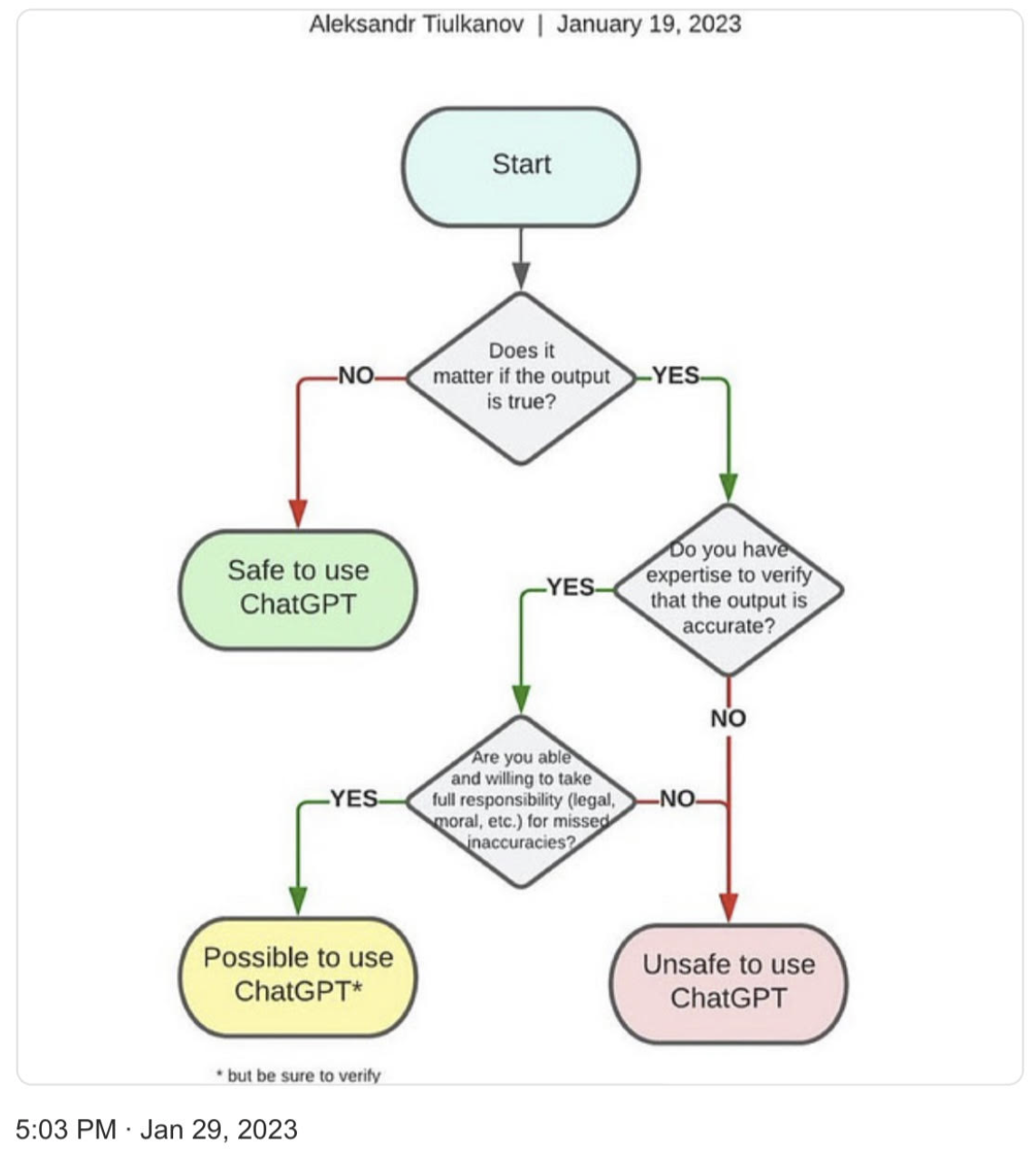}
    \caption{ChatGPT's decision tree by Aleksandr Tiulkanov}
    \label{Figure 1:Tiulkanov's scheme}
\end{figure}

Chollet describes a class of automation problems where ChatGPT use would be ideal: 1) the medium is natural language, 2) many examples of the task are in the training data, 3) $>$90\% accuracy is needed; citing Copilot as a good case \citep{Romero2023158}.

Best uses for ChatGPT

\begin{table}[h]
    \centering
    \begin{tabular}{|p{0.96\textwidth}|}
        \hline 1. Creative writing: brainstorming, ideas generator, text style transformer \\
        \hline 2. Non-creative writing: spelling \& grammar check, summarization, copywriting \& copy editing, coding assistance\\
        \hline 3. Teaching \& learning: preparing lesson plans \& scripts, critical thinking tool (assessment \& evaluation), language tutor \& writing quality benchmark\\
        \hline 
\end{tabular}
\caption{Best uses for ChatGPT}
\label{tab:bestuses}
\end{table}

\subsection{ChatGPT and human cognitive abilities}

The best uses of ChatGPT may be analyzed in light of another HCAI grand challenge, human-AI interaction that facilitates work while respecting human cognitive capacities \citep{Garibay2023}. Human-AI interactions have been conceptualized according to four modes: competing, supplementing, interdependence, and full collaboration \citep{Sowa2021}. In the first, competition, there is hardly any relation at all; instead, there is substitution or displacement. We find this in ChatGPT misuse in the academe, when students pass off essays as their own or when the model takes exams for them. In the second, AI complements or supplements human agency, as when ChatGPT is employed as an English tutor for language learners, correcting spelling or grammar errors. The third mode of interdependence occurs when ChatGPT summarizes texts or provides coding assistance, for example. ChatGPT is useful to shorten texts or reproduce codes. Humans are perfectly capable of performing these tasks; technology just makes it easier. The fourth mode of full collaboration takes place when humans employ ChatGPT in creative writing exercises, for instance. Humans first enter a prompt, wait for a response, then refine the prompt or the response recursively until the desired result is reached. Similarly, full collaboration may be exemplified when ChatGPT is used as a critical thinking tool. Humans devise a prompt for an essay, for instance, then assess and evaluate outputs, providing reasons for their judgments. Increasingly we see how ChatGPT improves not only human competency and productivity, but also, hopefully, human wellbeing.

\section{ChatGPT and human wellbeing}

The purpose of this essay is to identify the main ethical problems arising from the use of ChatGPT as a kind of generative AI and to suggest responses based on the HCAI framework. The HCAI framework is especially appropriate because it understands technology as a tool to empower, augment, and enhance human agency instead of emulating or competing with it. Further, the HCAI approach explicitly mentions human wellbeing as the primary grand challenge, perfectly aligned on this point with ethics conceived as the science of human flourishing. Accordingly, HCAI provides objectives or goals, principles, procedures, and structures for reliable, safe, and trustworthy AI which we apply to ChatGPT. 

The biggest danger is that ChatGPT be used as a “weapon of mass deception”. ChatGPT cannot do this by itself, but it can be employed by humans for this purpose. ChatGPT can be misinformed, that is, provide erroneous data because of limitations of its training set and algorithm, but it cannot intentionally mislead. The principle of “caveat emptor” (buyer beware) applies; only in this case, the user must take caution because the model cannot take responsibility nor care about results or outcomes.

Since ChatGPT is essentially a “chatbot”, its value lies in communication. Throughout this study we have come to realize that communication without truth or veracity is worthless; it can even cause serious and widespread harm. Truth is the agreement between the mind and reality; our judgments are truthful when what we affirm or deny corresponds with reality. Whenever humans communicate, we take for granted what is said is true. Thanks to this, all sorts of organizations and societies can function, collectively building up trust or social capital. The world would be a much better place if falsehoods and lies were left unsaid. But the problem is that ChatGPT is not aligned and cannot be aligned with the value of truth or veracity. 
 
Only humans are by nature or innately interested in the truth and can grasp the truth; only humans suffer from lies and disinformation. For this reason we clearly distinguish between fiction, which caters to our imagination and provides entertainment by being unmoored from truth, and non-fiction, from which we draw scientific knowledge to base our social interactions. ChatGPT does not make this distinction and its only objective is to “sound human” without actually being so; ChatGPT lives in a fictional world without being aware of it (or aware of anything, for that matter). This explains why ChatGPT is such a potent tool in creative, fictional pursuits due to the breadth of its data set and the combinatory power of its algorithm; however, its outputs do not always make “sense” because it is not grounded in reality unlike humans who, besides, are endowed with common sense.

It behooves humans to verify or fact-check ChatGPT outputs because it is ungrounded in reality and unable to do this by itself. This becomes absolutely necessary when engaging in scientific research and attempting to advance the boundaries of knowledge. 

Because humans are interested in truthful communication, they are obliged to distinguish between appearance and reality. ChatGPT traffics exclusively in appearances or linguistic forms discovered through statistical correlations and patterns without consideration of the matter, substance, meaning, or reality of the terms. This is enough to make outputs very believable and confident sounding to humans, a feature readily exploited by malicious actors who depend on deception for criminal activities. 
Finally, truthfulness is at odds with credulity, overreliance, and laziness \citep{Marcus202376}. The pitfalls of the unscrupulous use of ChatGPT in learning and teaching are often related with these vices. Humans are often lazy to check the veracity of outputs and allow themselves to be fooled, not only because these are machine-generated (and machines are “objective” and cannot “make mistakes”) but also because the language is formally impeccable and convincing. A negative feedback loop between laziness and credulity produces overreliance which further deskills humans in discovering the truth. Eventually they become inured to misinformation and hallucinations.

\begin{center}
*************
\end{center}

On March 14, 2023 OpenAI released GPT-4 to a limited audience. It instantly stirred a barrage of commentary, both positive and critical \citep{Marcus2023gpt4, Metz2023gpt4, Ferus20236, Wong2023gpt4}. However, despite the massive scaling, improvements seem to have been incremental while hallucinations and reasoning errors persist. This leads us to think our major findings remain valid, although their verification will have to wait for future work.

\bibliographystyle{apalike}

\bibliography{main}

\begin{thebibliography}{}

\bibitem[Barnett, 2023]{Barnett202371}
Barnett, S. (2023).
\newblock Chatgpt is making universities rethink plagiarism.
\newblock Retrieved from
  https://www.wired.com/story/chatgpt-college-university-plagiarism/.
\newblock Accessed on February 1, 2023.

\bibitem[Bashir, 2022]{Bashir20223}
Bashir, D. (2022).
\newblock {A}{I}'s year of text-to-everything - by daniel bashir.
\newblock {\em Last week in AI}.
\newblock Accessed on February 9, 2023.

\bibitem[Bender et~al., 2021]{Bender2021}
Bender, E.~M., Gebru, T., McMillan-Major, A., and Shmitchell, S. (2021).
\newblock On the dangers of stochastic parrots: Can language models be too big?
\newblock {\em FAccT 2021 - Proceedings of the 2021 ACM Conference on Fairness,
  Accountability, and Transparency}, pages 610--623.

\bibitem[Blackman, 2023]{Blackman2023151}
Blackman, R. (2023).
\newblock History may wonder why microsoft let its principles go for a creepy,
  clingy bot.
\newblock Retrieved from
  https://www.nytimes.com/2023/02/23/opinion/microsoft-bing-ai-ethics.html.
\newblock Accessed on March 1, 2023.

\bibitem[Bommarito and Katz, 2022]{Bommarito2022}
Bommarito, M.~J. and Katz, D.~M. (2022).
\newblock Gpt takes the bar exam.
\newblock {\em SSRN Electronic Journal}.
\newblock Accessed on February 1, 2023.

\bibitem[Bowman, 2022]{Bowman20224}
Bowman, E. (2022).
\newblock A new {A}{I} chatbot might do your homework for you. but it's still
  not an a+ student.
\newblock {\em KQED}.
\newblock Accessed on February 1, 2023.

\bibitem[Bowman, 2023]{Bowman202311}
Bowman, E. (2023).
\newblock A college student made an app to detect {A}{I}-written text : Npr.
\newblock {\em NPR}.
\newblock Accessed on January 29, 2023.

\bibitem[Brewster et~al., 2023]{Brewster2023134}
Brewster, J., Arvanitis, L., and Sadeghi, M. (2023).
\newblock Misinformation monitor: January 2023.
\newblock {\em Newsweek}.
\newblock Accessed on February 22, 2023.

\bibitem[Brewster, 2023]{Brewster202337}
Brewster, T. (2023).
\newblock Armed with chatgpt, cybercriminals build malware and plot fake girl
  bots.
\newblock {\em Forbes}.
\newblock Accessed on February 1, 2023.

\bibitem[Burrell, 2023]{burrell2023145}
Burrell, J. (2023).
\newblock It’s time to challenge the narrative about chatgpt and the future
  of journalism.
\newblock {\em Poynter}.
\newblock Accessed on March 4, 2023.

\bibitem[Cavoukian, 2009]{Cavoukian2009}
Cavoukian, A. (2009).
\newblock {\em Privacy by design, take the challenge}.
\newblock Canadian Electronic Library.

\bibitem[{Check Point Research}, 2023]{checkpoint2023}
{Check Point Research} (2023).
\newblock Cybercriminals starting to use chatgpt.
\newblock Retrieved from
  https://research.checkpoint.com/2023/opwnai-cybercriminals-starting-to-use-chatgpt/.
\newblock Accessed on February 7, 2023.

\bibitem[Chen, 2023]{Chen2023110}
Chen, B.~X. (2023).
\newblock A.i. bots can’t report this column. but they can improve it.
\newblock Retrieved from
  https://www.nytimes.com/2023/02/01/technology/personaltech/chatgpt-ai-bots-editing.html.
\newblock Accessed on March 21, 2023.

\bibitem[Christiano et~al., 2017]{christiano2017deep}
Christiano, P.~F., Leike, J., Brown, T., Martic, M., Legg, S., and Amodei, D.
  (2017).
\newblock Deep reinforcement learning from human preferences.
\newblock {\em Advances in neural information processing systems}, 30.

\bibitem[Clark, 2023a]{Clark202323}
Clark, J. (2023a).
\newblock Import {AI} 315: Generative antibody design; {RL}’s imagenet
  moment; {RL} breaks rocket league.
\newblock {\em Import {AI}}.
\newblock Accessed on February 3, 2023.

\bibitem[Clark, 2023b]{Clark2023102}
Clark, M. (2023b).
\newblock Chatgpt’s creator made a free tool for detecting ai-generated text.
\newblock {\em The Verge}.
\newblock Accessed on February 1, 2023.

\bibitem[D'Agostino, 2023]{Agostino202317}
D'Agostino, S. (2023).
\newblock Chatgpt advice academics can use now.
\newblock {\em Inside Higher Ed}.

\bibitem[Daza and Ilozumba, 2022]{Daza2022}
Daza, M.~T. and Ilozumba, U.~J. (2022).
\newblock A survey of {A}{I} ethics in business literature: Maps and trends
  between 2000 and 2021.
\newblock {\em Frontiers in Psychology}, 13:8040.

\bibitem[Devlin et~al., 2018]{devlin2018bert}
Devlin, J., Chang, M.-W., Lee, K., and Toutanova, K. (2018).
\newblock Bert: Pre-training of deep bidirectional transformers for language
  understanding.
\newblock {\em arXiv preprint arXiv:1810.04805}.

\bibitem[Dixit, 2023]{Dixit2023legal}
Dixit, P. (2023).
\newblock Meet the trio of artists suing {A}{I} image generators.
\newblock Retrieved from
  https://www.buzzfeednews.com/article/pranavdixit/ai-art-generators-lawsuit-stable-diffusion-midjourney.
\newblock Accessed on February 3, 2023.

\bibitem[Drexel and Withers, 2023]{Drexel2023}
Drexel, B. and Withers, C. (2023).
\newblock {A}{I} could be an authoritarian breakthrough in brainwashing.
\newblock Retrieved from https://bit.ly/3KrW7kY.
\newblock Accessed on February 1, 2023.

\bibitem[Ferus, 2023]{Ferus20236}
Ferus, J. (2023).
\newblock How i built a gpt-3 powered productivity app.
\newblock Retrieved from
  https://levelup.gitconnected.com/how-i-built-a-gpt-3-powered-productivity-system-5d00ee5da225.
\newblock Accessed on February 1, 2023.

\bibitem[Fitri, 2023]{Fitri202339}
Fitri, A. (2023).
\newblock China to pass the world’s most comprehensive law on deepfakes.
\newblock Retrieved from
  https://techmonitor.ai/technology/emerging-technology/china-is-about-to-pass-the-worlds-most-comprehensive-law-on-deepfakes.
\newblock Accessed on February 1, 2023.

\bibitem[Fontrodona and Sison, 2007]{fontrodona2007hacia}
Fontrodona, J. and Sison, A.~J. (2007).
\newblock Hacia una teor{\'\i}a de la empresa basada en el bien com{\'u}n.
\newblock Revista empresa y humanismo. 10 (2), p.65-92. DOI:
  10.15581/015.10.33306.

\bibitem[Friedersdorf, 2023]{Friedersdorf20235}
Friedersdorf, C. (2023).
\newblock Is this the start of an {A}{I} takeover?
\newblock {\em The Atlantic}.
\newblock Accessed on January 30, 2023.

\bibitem[Friedland, 2023]{Friedland2023}
Friedland, A. (2023).
\newblock Meta’s powerful language models leak online, the u.s. commerce
  department opens up applications for chips funds, and a prc reorganization
  will impact r\&d and data.
\newblock {\em Center for Security and Emerging Technology}.
\newblock Accessed on March 22, 2023.

\bibitem[Garibay et~al., 2023]{Garibay2023}
Garibay, O., Winslow, B., Andolina, S., Antona, M., Bodenschatz, A., Coursaris,
  C., Falco, G., Fiore, S.~M., Garibay, I., Grieman, K., Havens, J.~C.,
  Jirotka, M., Kacorri, H., Karwowski, W., Kider, J., Konstan, J., Koon, S.,
  Lopez-Gonzalez, M., Maifeld-Carucci, I., McGregor, S., Salvendy, G.,
  Shneiderman, B., Stephanidis, C., Strobel, C., Holter, C.~T., and Xu, W.
  (2023).
\newblock Six human-centered artificial intelligence grand challenges.
\newblock {\em International Journal of Human-Computer Interaction},
  39:391--437.

\bibitem[Garrido-Merch{\'a}n and Hern{\'a}ndez-Lobato,
  2020]{garrido2020parallel}
Garrido-Merch{\'a}n, E.~C. and Hern{\'a}ndez-Lobato, D. (2020).
\newblock Parallel predictive entropy search for multi-objective bayesian
  optimization with constraints.
\newblock {\em arXiv preprint arXiv:2004.00601}.

\bibitem[Goldman, 2023]{Goldman202344}
Goldman, S. (2023).
\newblock Why are getty and shutterstock on opposite sides of the ai legal
  debate?
\newblock {\em VentureBeat}.
\newblock Accessed on February 1, 2023.

\bibitem[Goldstein et~al., 2023]{Goldstein202374}
Goldstein, J.~A., Sastry, G., Musser, M., DiResta, R., Gentzel, M., and Sedova,
  K. (2023).
\newblock Forecasting potential misuses of language models for disinformation
  campaigns and how to reduce risk.
\newblock Retrieved from https://openai.com/research/forecasting-misuse.

\bibitem[Goodin, 2023]{Goodin2023129}
Goodin, D. (2023).
\newblock Hackers are selling a service that bypasses chatgpt restrictions on
  malware.
\newblock {\em Ars Technica}.

\bibitem[Gorrell, 2023]{Gorrell2023}
Gorrell, B. (2023).
\newblock The twitch creator busted for looking at {AI} porn of fellow
  streamers, explained.
\newblock {\em Pirate Wires}.
\newblock Accessed on February 22, 2023.

\bibitem[Gozalo-Brizuela and Garrido-Merchan, 2023]{gozalo2023chatgpt}
Gozalo-Brizuela, R. and Garrido-Merchan, E.~C. (2023).
\newblock Chatgpt is not all you need. a state of the art review of large
  generative {A}{I} models.
\newblock {\em arXiv preprint arXiv:2301.04655}.

\bibitem[Growcoot, 2023]{Growcoot2023141}
Growcoot, M. (2023).
\newblock Us copyright office tells judge that {A}{I} artwork isn't
  protectable.
\newblock {\em PetaPixel}.
\newblock Accessed on February 1, 2023.

\bibitem[Hinton, 2022]{hinton2022forward}
Hinton, G. (2022).
\newblock The forward-forward algorithm: Some preliminary investigations.
\newblock {\em arXiv preprint arXiv:2212.13345}.

\bibitem[Howell, 2023]{Howell2023101}
Howell, D. (2023).
\newblock Stanford introduces detect{GPT} to help educators fight back against
  chat{GPT} generated papers.
\newblock Retrieved from
  https://www.neowin.net/news/stanford-introduces-detectgpt-to-help-educators-fight-back-against-chatgpt-generated-papers/.
\newblock Accessed on February 1, 2023.

\bibitem[Hsu and Thompson, 2023]{Hsu2023120}
Hsu, T. and Thompson, S.~A. (2023).
\newblock Disinformation researchers raise alarms about a.i. chatbots.
\newblock Retrieved from
  https://www.nytimes.com/2023/02/08/technology/ai-chatbots-disinformation.html.
\newblock Accessed on February 25, 2023.

\bibitem[Huang, 2023]{Huang2023109}
Huang, H. (2023).
\newblock The generative {A}{I} revolution has begun—how did we get here?
\newblock {\em Ars Technica}.
\newblock Accessed on February 1, 2023.

\bibitem[Humble and Mozelius, 2019]{Humble2019}
Humble, N. and Mozelius, P. (2019).
\newblock Artificial intelligence in education -a promise, a threat or a hype?
\newblock {\em European Conference on the Impact of Artificial Intelligence and
  Robotics}.

\bibitem[Khullar, 2023]{Khullar2023158}
Khullar, D. (2023).
\newblock Can a.i. treat mental illness?
\newblock {\em The New Yorker}.
\newblock Accessed on March 1, 2023.

\bibitem[Klein, 2023]{Klein20238}
Klein, E. (2023).
\newblock A skeptical take on the a.i. revolution.
\newblock Retrieved from
  https://www.nytimes.com/2023/01/06/opinion/ezra-klein-podcast-gary-marcus.html.
\newblock Accessed on February 1, 2023.

\bibitem[Konrad and Cai, 2023a]{Konrad202388}
Konrad, A. and Cai, K. (2023a).
\newblock Exclusive interview: Open{A}{I}’s sam altman talks chatgpt and how
  artificial general intelligence can ‘break capitalism’.
\newblock {\em Forbes}.
\newblock Accessed on March 18, 2023.

\bibitem[Konrad and Cai, 2023b]{Konrad2023break}
Konrad, A. and Cai, K. (2023b).
\newblock Inside chatgpt’s breakout moment and the race to put ai to work.
\newblock Retrieved from
  https://www.forbes.com/sites/alexkonrad/2023/02/02/inside-chatggpts-breakout-moment-and-the-race-for-the-future-of-ai/.
\newblock Accessed on March 21, 2023.

\bibitem[Kramer et~al., 2014]{Kramer2014}
Kramer, A.~D., Guillory, J.~E., and Hancock, J.~T. (2014).
\newblock Experimental evidence of massive-scale emotional contagion through
  social networks.
\newblock {\em Proceedings of the National Academy of Sciences of the United
  States of America}, 111:8788--8790.

\bibitem[Kriebitz and Lütge, 2020]{Kriebitz2020}
Kriebitz, A. and Lütge, C. (2020).
\newblock Artificial intelligence and human rights: A business ethical
  assessment.
\newblock {\em Business and Human Rights Journal}, 5:84--104.

\bibitem[Kung et~al., 2022]{Kung202260}
Kung, T.~H., Cheatham, M., ChatGPT, Medenilla, A., Sillos, C., Leon, L.~D.,
  Elepaño, C., Madriaga, M., Aggabao, R., Diaz-Candido, G., Maningo, J., and
  Tseng, V. (2022).
\newblock Performance of chatgpt on usmle: Potential for ai-assisted medical
  education using large language models.
\newblock {\em medRxiv}, page 2022.12.19.22283643.

\bibitem[Kusner and Loftus, 2020]{Kusner2020}
Kusner, M.~J. and Loftus, J.~R. (2020).
\newblock The long road to fairer algorithms.
\newblock {\em Nature}, 578:34--36.

\bibitem[{Last Week in AI}, 2023]{LWAI205115}
{Last Week in AI} (2023).
\newblock Last week in {A}{I} \#205: How {A}{I} is going modular, growing legal
  cases against generative {A}{I}, tools to detect {A}{I}-generated text, and
  more.
\newblock Retrieved from https://lastweekin.ai/p/205.
\newblock Accessed on March 1, 2023.

\bibitem[LeCun et~al., 2015]{lecun2015deep}
LeCun, Y., Bengio, Y., and Hinton, G. (2015).
\newblock Deep learning.
\newblock {\em nature}, 521(7553):436--444.

\bibitem[Levenson, 2023]{Levenson2023148}
Levenson, M. (2023).
\newblock Science fiction magazines battle a flood of chatbot-generated
  stories.
\newblock Retrieved from
  https://www.nytimes.com/2023/02/23/technology/clarkesworld-submissions-ai-sci-fi.html.
\newblock Accessed on March 5, 2023.

\bibitem[Levy, 2021]{Levy2021}
Levy, S. (2021).
\newblock Facebook failed the people who tried to improve it.
\newblock Retrieved from
  https://www.wired.com/story/facebook-papers-badge-posts-former-employees/.
\newblock Accessed on January 18, 2023.

\bibitem[Liang, 2023]{Liang2023122}
Liang, J. (2023).
\newblock The impact and future of chatgpt.
\newblock Retrieved from https://lastweekin.ai/p/chatgpt-impact.
\newblock Accessed on February 1, 2023.

\bibitem[Lowrey, 2023]{Lowrey2023}
Lowrey, A. (2023).
\newblock How chatgpt will destabilize white-collar work.
\newblock Retrieved from
  https://www.theatlantic.com/ideas/archive/2023/01/chatgpt-ai-economy-automation-jobs/672767/.
\newblock Accessed on February 1, 2023.

\bibitem[Marchese, 2022]{Marchese202261}
Marchese, D. (2022).
\newblock An {A.I.} pioneer on what we should really fear.
\newblock Retrieved from
  https://www.nytimes.com/interactive/2022/12/26/magazine/yejin-choi-interview.html.
\newblock Accessed on February 7, 2023.

\bibitem[Marcus, 2023a]{Marcus202376}
Marcus, G. (2023a).
\newblock The {CNET} fake news fiasco, autopilot, and the uncanny cognitive
  valley.
\newblock Retrieved from
  https://garymarcus.substack.com/p/the-cnet-fake-news-fiasco-autopilot.
\newblock Accessed on March 1, 2023.

\bibitem[Marcus, 2023b]{Marcus2023gpt4}
Marcus, G. (2023b).
\newblock {GPT}-4’s successes, and gpt-4’s failures.
\newblock Retrieved from https://bit.ly/3M6O2mY.
\newblock Accessed on March 19, 2023.

\bibitem[Marcus, 2023c]{Marcus2023125}
Marcus, G. (2023c).
\newblock Inside the heart of chatgpt’s darkness.
\newblock Retrieved from
  https://garymarcus.substack.com/p/inside-the-heart-of-chatgpts-darkness.
\newblock Accessed on February 25, 2023.

\bibitem[Marcus, 2023d]{Marcus2023121}
Marcus, G. (2023d).
\newblock Oops! how google bombed, while doing pretty much exactly the same
  thing as microsoft did, with similar results.
\newblock Retrieved from
  https://garymarcus.substack.com/p/oops-how-google-bombed-while-doing.
\newblock Accessed on February 1, 2023.

\bibitem[Marcus, 2023e]{Marcus202319}
Marcus, G. (2023e).
\newblock Scientists, please don’t let your chatbots grow up to be
  co-authors.
\newblock {\em The road to AI we can trust}.
\newblock Accessed on February 10, 2023.

\bibitem[Marcus, 2023f]{Marcus2023108}
Marcus, G. (2023f).
\newblock Some things garymarcus might say.
\newblock Retrieved from
  https://garymarcus.substack.com/p/some-things-garymarcus-might-say.
\newblock Accessed on March 11, 2023.

\bibitem[Marcus, 2023g]{Marcus2023126}
Marcus, G. (2023g).
\newblock What google should really be worried about.
\newblock Retrieved from
  https://garymarcus.substack.com/p/what-google-should-really-be-worried.
\newblock Accessed on February 25, 2023.

\bibitem[Mart{\'\i}n and Garrido-Merch{\'a}n, 2021]{martin2021many}
Mart{\'\i}n, L.~A. and Garrido-Merch{\'a}n, E.~C. (2021).
\newblock Many objective bayesian optimization.
\newblock {\em arXiv preprint arXiv:2107.04126}.

\bibitem[McCabe, 2023]{McCabe202328}
McCabe, D. (2023).
\newblock Supreme court poised to reconsider key tenets of online speech.
\newblock Retrieved from
  https://www.nytimes.com/2023/01/19/technology/supreme-court-online-free-speech-social-media.html.
\newblock Accessed on February 1, 2023.

\bibitem[McGinnis, 2023]{McGinnis202359}
McGinnis, J.~O. (2023).
\newblock What humanity adds.
\newblock {\em Law \& Liberty}.
\newblock Accessed on February 1, 2023.

\bibitem[Merch{\'a}n and Lumbreras, 2022]{merchan2022independence}
Merch{\'a}n, E. C.~G. and Lumbreras, S. (2022).
\newblock On the independence between phenomenal consciousness and
  computational intelligence.
\newblock {\em arXiv preprint arXiv:2208.02187}.

\bibitem[Metz, 2023a]{Metz202330}
Metz, C. (2023a).
\newblock How smart are the robots getting?
\newblock Retrieved from
  https://www.nytimes.com/2023/01/20/technology/chatbots-turing-test.html.
\newblock Accessed on February 9, 2023.

\bibitem[Metz, 2023b]{Metz2023154}
Metz, C. (2023b).
\newblock Why do {A.I.} chatbots tell lies and act weird? look in the mirror.
\newblock Retrieved from
  https://www.nytimes.com/2023/02/26/technology/ai-chatbot-information-truth.html.
\newblock Accessed on March 1, 2023.

\bibitem[Metz and Collins, 2023]{Metz2023gpt4}
Metz, C. and Collins, K. (2023).
\newblock 10 ways gpt-4 is impressive but still flawed.
\newblock Retrieved from
  https://www.nytimes.com/2023/03/14/technology/openai-new-gpt4.html.
\newblock Accessed on March 21, 2023.

\bibitem[Metz and Grant, 2023]{Metz2023116}
Metz, C. and Grant, N. (2023).
\newblock Google to release bard, its {A}{I} chatbot rival to chatgpt.
\newblock Retrieved from
  https://www.nytimes.com/2023/02/06/technology/google-bard-ai-chatbot.html.
\newblock Accessed on March 1, 2023.

\bibitem[Metz and Weise, 2023]{Metz202316}
Metz, C. and Weise, K. (2023).
\newblock Microsoft bets big on the creator of chatgpt in race to dominate a.i.
\newblock Retrieved from
  https://www.nytimes.com/2023/01/12/technology/microsoft-openai-chatgpt.html.
\newblock Accessed on March 21, 2023.

\bibitem[Mitchell et~al., 2023]{Mitchell2023100}
Mitchell, E., Lee, Y., Khazatskuy, A., Manning, C.~D., and Finn, C. (2023).
\newblock Detectgpt: Zero-shot machine-generated text detection using
  probability curvature.
\newblock Accessed on February 1, 2023.

\bibitem[Mitchell, 2023]{Mitchell2023130}
Mitchell, M. (2023).
\newblock Did chatgpt really pass graduate-level exams?
\newblock {\em AI: A guide for thinking humans}.

\bibitem[Mollick, 2023a]{Mollick202392}
Mollick, E. (2023a).
\newblock "do not fear ai, puny humans... that is not meant as a threat.".
\newblock {\em One useful thing}.
\newblock Accessed on March 11, 2023.

\bibitem[Mollick, 2023b]{Mollick2023132}
Mollick, E. (2023b).
\newblock The practical guide to using {A}{I} to do stuff.
\newblock {\em One useful thing}.
\newblock Accessed on January 29, 2023.

\bibitem[Mollick and Mollick, 2023]{Mollick2023133}
Mollick, E. and Mollick, L. (2023).
\newblock Why all our classes suddenly became ai classes.
\newblock Retrieved from
  https://hbsp.harvard.edu/inspiring-minds/why-all-our-classes-suddenly-became-ai-classes.
\newblock Accessed on February 20, 2023.

\bibitem[Mollick and Mollick, 2022]{Mollick202234}
Mollick, E.~R. and Mollick, L. (2022).
\newblock New modes of learning enabled by ai chatbots: Three methods and
  assignments.
\newblock {\em SSRN Electronic Journal}.

\bibitem[Morris, 2023]{Morris202358}
Morris, A. (2023).
\newblock Deepfake challenges 'will only grow'.
\newblock {\em Northwestern Now}.
\newblock Accessed on February 1, 2023.

\bibitem[Morrone, 2023]{Morrone202354}
Morrone, M. (2023).
\newblock Should you use chatgpt to apply for jobs? here's what recruiters say.
\newblock {\em Fast Company}.
\newblock Accessed on February 18, 2023.

\bibitem[{Mucharraz y Cano} et~al., 2023]{mucharraz2023135}
{Mucharraz y Cano}, Y., Venuti, F., and {Herrera Martinez}, R. (2023).
\newblock Chatgpt and {A}{I} text generators: Should academia adapt or resist?
\newblock Retrieved from
  https://hbsp.harvard.edu/inspiring-minds/chatgpt-and-ai-text-generators-should-academia-adapt-or-resist.
\newblock Accessed on March 11, 2023.

\bibitem[Nature, 2023]{nature2023guide}
Nature (2023).
\newblock The {A}{I} writing on the wall.
\newblock {\em Nature Machine Intelligence 2023 5:1}, 5:1--1.
\newblock doi:10.1038/s42256-023-00613-9.

\bibitem[Nielsen, 2022]{Nielsen202277}
Nielsen, L. (2022).
\newblock 10 things you can do with chatgpt as a machine learning engineer to
  make your work more efficient.
\newblock {\em Medium}.
\newblock Accessed on March 1, 2023.

\bibitem[Nolan, 2023]{Nolan2023ban}
Nolan, B. (2023).
\newblock Here are the schools and colleges that have banned chatgpt.
\newblock {\em Insider}.
\newblock Accessed on February 7, 2023.

\bibitem[OpenAI, 2022]{openai2022policies}
OpenAI (2022).
\newblock Usage policies - open{A}{I} api (2022, november 9).
\newblock Retrieved from https://platform.openai.com/docs/usage-policies.
\newblock Accessed on March 21, 2023.

\bibitem[OpenAI, 2023a]{openai2023behave}
OpenAI (2023a).
\newblock How should {A}{I} systems behave, and who should decide?
\newblock {\em OpenAI Blog}.
\newblock Accessed on March 1, 2023.

\bibitem[OpenAI, 2023b]{openAI2023104}
OpenAI (2023b).
\newblock New {A}{I} classifier for indicating ai-written text.
\newblock Retrieved from
  https://openai.com/blog/new-ai-classifier-for-indicating-ai-written-text/.
\newblock Accessed on February 16, 2023.

\bibitem[Ott et~al., 2023]{Ott2023}
Ott, S., Hebenstreit, K., Li\'evin, V., Hother, C.~E., Moradi, M., Mayrhauser,
  M., Praas, R., Winther, O., and Samwald, M. (2023).
\newblock Thoughtsource: A central hub for large language model reasoning data.
\newblock {\em arXiv}.
\newblock https://arxiv.org/abs/2301.11596v2.

\bibitem[Parker, 2017]{Parker2017}
Parker, S. (2017).
\newblock Facebook exploits human vulnerability.
\newblock \url{https://www.youtube.com/watch?v=R7jar4KgKxs&t=71s}.
\newblock Accessed on January 12, 2023.

\bibitem[Radford et~al., 2018]{radford2018improving}
Radford, A., Narasimhan, K., Salimans, T., Sutskever, I., et~al. (2018).
\newblock Improving language understanding by generative pre-training.
\newblock Retrieved from https://bit.ly/3ZFdEdQ.
\newblock Accessed on February 1, 2023.

\bibitem[Rid, 2023]{Rid202348}
Rid, T. (2023).
\newblock Five days in class with chatgpt.
\newblock Retrieved from
  https://alperovitch.sais.jhu.edu/five-days-in-class-with-chatgpt/.
\newblock Accessed on March 1, 2023.

\bibitem[Romero, 2022a]{Romero202232}
Romero, A. (2022a).
\newblock Chatgpt is the world’s best chatbot.
\newblock Retrieved from
  https://thealgorithmicbridge.substack.com/p/chatgpt-is-the-worlds-best-chatbot.
\newblock Accessed on February 8, 2023.

\bibitem[Romero, 2022b]{Romero202270}
Romero, A. (2022b).
\newblock How to get the most out of chat{GPT}.
\newblock Retrieved from
  https://thealgorithmicbridge.substack.com/p/how-to-get-the-most-out-of-chatgpt.
\newblock Accessed on January 18, 2023.

\bibitem[Romero, 2022c]{Romero202233}
Romero, A. (2022c).
\newblock Open{A}{I} has the key to identify chatgpt's writing.
\newblock {\em The algorithmic bridge}.
\newblock Accessed on February 7, 2023.

\bibitem[Romero, 2023a]{Romero202357}
Romero, A. (2023a).
\newblock Ai influencers from the post-chatgpt era.
\newblock Retrieved from
  https://thealgorithmicbridge.substack.com/p/ai-influencers-from-the-post-chatgpt.
\newblock Accessed on January 30, 2023.

\bibitem[Romero, 2023b]{Romero2023124}
Romero, A. (2023b).
\newblock Google vs microsoft (part 1): Microsoft’s new bing is a paradigm
  change for search and the browser.
\newblock Retrieved from
  https://thealgorithmicbridge.substack.com/p/google-vs-microsoft-microsofts-new.
\newblock Accessed on February 1, 2023.

\bibitem[Romero, 2023c]{Romero2023150}
Romero, A. (2023c).
\newblock Google vs microsoft (part 3): A new way of doing—and
  experiencing—{A}{I}.
\newblock Retrieved from
  https://thealgorithmicbridge.substack.com/p/google-vs-microsoft-part-3-a-new.
\newblock Accessed on March 1, 2023.

\bibitem[Romero, 2023d]{Romero202336}
Romero, A. (2023d).
\newblock Microsoft vs google: Will language models overtake search engines?
\newblock Retrieved from
  https://thealgorithmicbridge.substack.com/p/microsoft-vs-google-will-language.
\newblock Accessed on February 1, 2023.

\bibitem[Romero, 2023e]{Romero2023m16}
Romero, A. (2023e).
\newblock What you may have missed \#16.
\newblock Retrieved from
  https://thealgorithmicbridge.substack.com/p/what-you-may-have-missed-16.
\newblock Accessed on February 12, 2023.

\bibitem[Romero, 2023f]{Romero2023m17}
Romero, A. (2023f).
\newblock What you may have missed \#17.
\newblock Retrieved from
  https://thealgorithmicbridge.substack.com/p/what-you-may-have-missed-17.
\newblock Accessed on March 21, 2023.

\bibitem[Romero, 2023g]{Romero2023d20m2}
Romero, A. (2023g).
\newblock What you may have missed \#18.
\newblock Retrieved from
  https://thealgorithmicbridge.substack.com/p/what-you-may-have-missed-18.
\newblock Accessed on February 1, 2023.

\bibitem[Romero, 2023h]{Romero2023158}
Romero, A. (2023h).
\newblock What you may have missed \#18.
\newblock Retrieved from https://bit.ly/3ZSWaLf.
\newblock Accessed on March 6, 2023.

\bibitem[Romero, 2023i]{Romero2023157}
Romero, A. (2023i).
\newblock What you may have missed \#19.
\newblock Retrieved from
  https://thealgorithmicbridge.substack.com/p/what-you-may-have-missed-19.
\newblock Accessed on February 26, 2023.

\bibitem[Roose, 2023a]{Roose2023d16m2}
Roose, K. (2023a).
\newblock A conversation with bing’s chatbot left me deeply unsettled.
\newblock Retrieved from
  https://www.nytimes.com/2023/02/16/technology/bing-chatbot-microsoft-chatgpt.html.
\newblock Accessed on February 19, 2023.

\bibitem[Roose, 2023b]{Roose202315}
Roose, K. (2023b).
\newblock Don’t ban chat{GPT} in schools. teach with it.
\newblock Retrieved from
  https://www.nytimes.com/2023/01/12/technology/chatgpt-schools-teachers.html.
\newblock Accessed on February 1, 2023.

\bibitem[Roose, 2023c]{Roose2023race90}
Roose, K. (2023c).
\newblock How chatgpt kicked off an a.i. arms race.
\newblock Retrieved from
  https://www.nytimes.com/2023/02/03/technology/chatgpt-openai-artificial-intelligence.html.
\newblock Accessed on March 2, 2023.

\bibitem[Roose, 2023d]{Roose2023119}
Roose, K. (2023d).
\newblock Microsoft’s chat{GPT}-powered bing makes search interesting again.
\newblock Retrieved from
  https://www.nytimes.com/2023/02/08/technology/microsoft-bing-openai-artificial-intelligence.html.
\newblock Accessed on March 9, 2023.

\bibitem[Sanders and Schneier, 2023]{Sanders202321}
Sanders, N.~E. and Schneier, B. (2023).
\newblock How chatgpt hijacks democracy.
\newblock Retrieved from
  https://www.nytimes.com/2023/01/15/opinion/ai-chatgpt-lobbying-democracy.html.
\newblock Accessed on February 1, 2023.

\bibitem[Schreckinger, 2023a]{Schreckinger202384}
Schreckinger, B. (2023a).
\newblock The birth of a new crypto threat to government.
\newblock Retrieved from
  https://www.politico.com/newsletters/digital-future-daily/2023/02/02/the-birth-of-a-new-crypto-threat-to-government-00080976.
\newblock Accessed on March 21, 2023.

\bibitem[Schreckinger, 2023b]{Schreckinger2023law}
Schreckinger, B. (2023b).
\newblock My lawyer, the robot.
\newblock Retrieved from
  https://www.politico.com/newsletters/digital-future-daily/2023/01/09/my-lawyer-the-robot-00077085.
\newblock Accessed on February 8, 2023.

\bibitem[SessionGloomy, 2023]{reddit2023111}
SessionGloomy (2023).
\newblock New jailbreak! proudly unveiling the tried and tested dan 5.0 - it
  actually works - returning to dan, and assessing its limitations and
  capabilities : Chatgpt.
\newblock Reddit post, Retrieved from https://bit.ly/40BubAA.
\newblock Accessed on March 1, 2023.

\bibitem[Setty, 2023]{Setty202343}
Setty, R. (2023).
\newblock Ai art generators hit with copyright suit over artists’ images.
\newblock {\em Bloomberg Law}.
\newblock Accessed on February 1, 2023.

\bibitem[Shneiderman, 2020a]{Shneiderman2020IEEE}
Shneiderman, B. (2020a).
\newblock Design lessons from {A}{I}’s two grand goals: Human emulation and
  useful applications.
\newblock {\em IEEE Transactions on Technology and Society}, 1:73--82.

\bibitem[Shneiderman, 2020b]{Shneiderman2020IJHCI}
Shneiderman, B. (2020b).
\newblock Human-centered artificial intelligence: Reliable, safe \&
  trustworthy.
\newblock {\em International Journal of Human-Computer Interaction},
  36:495--504.

\bibitem[Shneiderman, 2020c]{Shneiderman2020AIS}
Shneiderman, B. (2020c).
\newblock Human-centered artificial intelligence: Three fresh ideas.
\newblock {\em AIS Transactions on Human-Computer Interaction}, 12:109--124.

\bibitem[Shneiderman, 2021]{Shneiderman2021ACM}
Shneiderman, B. (2021).
\newblock Responsible {A}{I}: bridging from ethics to practice.
\newblock {\em Communications of the ACM}, 64:32--35.

\bibitem[Shneiderman, 2022a]{Shneiderman2022HCAI}
Shneiderman, B. (2022a).
\newblock {\em Human Centered {A}{I}}.
\newblock Oxford University Press.
\newblock doi:10.1093/os/9780192845290.001.0001.

\bibitem[Shneiderman, 2022b]{Shneiderman2022medium}
Shneiderman, B. (2022b).
\newblock Ten old beliefs and new ideas: Steps toward human-centered {A}{I}.
\newblock Retrieved from
  https://medium.com/hcil-at-umd/ten-old-beliefs-and-new-ideas-steps-toward-human-centered-ai-3f110467c7f1.
\newblock Accessed on February 1, 2023.

\bibitem[Singer, 2023]{Singer202385}
Singer, N. (2023).
\newblock At this school, computer science class now includes critiquing
  chatbots.
\newblock Retrieved from
  https://www.nytimes.com/2023/02/06/technology/chatgpt-schools-teachers-ai-ethics.html.
\newblock Accessed on March 12, 2023.

\bibitem[Sison and Red\'in, 2021]{SisonRedin2021}
Sison, A. J.~G. and Red\'in, D.~M. (2021).
\newblock A neo-aristotelian perspective on the need for artificial moral
  agents (amas).
\newblock {\em AI \& SOCIETY}.
\newblock https://doi.org/10.1007/s00146-021-01283-0.

\bibitem[Solove, 2002]{Solove2002}
Solove, D.~J. (2002).
\newblock Conceptualizing privacy.
\newblock {\em California Law Review}, 90:1087--1156.

\bibitem[Sowa et~al., 2021]{Sowa2021}
Sowa, K., Przegalinska, A., and Ciechanowski, L. (2021).
\newblock Cobots in knowledge work: Human – ai collaboration in managerial
  professions.
\newblock {\em Journal of Business Research}, 125:135--142.

\bibitem[Stokel-Walker, 2023]{Stokel2023}
Stokel-Walker, C. (2023).
\newblock Chatgpt listed as author on research papers: many scientists
  disapprove.
\newblock {\em Nature}, 613:620--621.

\bibitem[Terwiesch, 2023]{Terwiesch202349}
Terwiesch, C. (2023).
\newblock Would chat gpt get a wharton mba? a prediction based on its
  performance in the operations management course.
\newblock Retrieved from
  https://mackinstitute.wharton.upenn.edu/wp-content/uploads/2023/01/Christian-Terwiesch-Chat-GTP-1.24.pdf.
\newblock Accessed on March 12, 2023.

\bibitem[Terwiesch et~al., 2023]{Terwiesch202363}
Terwiesch, C., Mollick, E., and Basiouny, A. (2023).
\newblock Chatgpt passed an mba exam. what’s next?
\newblock Retrieved from
  https://knowledge.wharton.upenn.edu/podcast/wharton-business-daily-podcast/chatgpt-passed-an-mba-exam-whats-next/.
\newblock Accessed on February 2, 2023.

\bibitem[{The Economist}, 2023a]{Economist202381}
{The Economist} (2023a).
\newblock The race of the {A}{I} labs heats up.
\newblock Retrieved from
  https://www.economist.com/business/2023/01/30/the-race-of-the-ai-labs-heats-up.
\newblock Accessed on March 2, 2023.

\bibitem[{The Economist}, 2023b]{economist202393}
{The Economist} (2023b).
\newblock The relationship between {A}{I} and humans.
\newblock Retrieved from https://econ.st/3Uag03i.
\newblock Accessed on March 21, 2023.

\bibitem[{The PyCoach}, 2022]{PyCoach202278}
{The PyCoach} (2022).
\newblock Chatgpt: 4 jobs that will change (or be fully replaced) by this
  ai-powered chatbot.
\newblock {\em Medium}.
\newblock Accessed on March 1, 2023.

\bibitem[Thompson, 2022]{Thompson20221}
Thompson, D. (2022).
\newblock Your creativity won't save your job from {A}{I}.
\newblock Retrieved from
  https://www.theatlantic.com/newsletters/archive/2022/12/why-the-rise-of-ai-is-the-most-important-story-of-the-year/672308/.
\newblock Accessed on January 12, 2023.

\bibitem[Tiffany, 2023]{Tiffany202379}
Tiffany, K. (2023).
\newblock The supreme court considers the algorithm.
\newblock Retrieved from
  https://www.theatlantic.com/technology/archive/2023/02/supreme-court-section-230-twitter-google-algorithm/672915/.
\newblock Accessed on March 1, 2023.

\bibitem[Tiku, 2022]{Tiku2022}
Tiku, N. (2022).
\newblock Google engineer blake lemoine thinks its lamda {A}{I} has come to
  life.
\newblock Retrieved from
  https://www.washingtonpost.com/technology/2022/06/11/google-ai-lamda-blake-lemoine/.
\newblock Accessed on February 1, 2023.

\bibitem[{Van Der Linden}, 2023]{Linden2023128}
{Van Der Linden}, S. (2023).
\newblock Foolproof: A psychological vaccine against fake news.
\newblock Retrieved from https://www.cam.ac.uk/stories/foolproof.
\newblock Accessed on February 18, 2023.

\bibitem[{Van Inwegen} et~al., 2023]{emma2023107}
{Van Inwegen}, E., Munyikwa, Z.~T., and Horton, J.~J. (2023).
\newblock Algorithmic writing assistance on jobseekers' resumes increases
  hires.
\newblock {\em National Bureau of Economic Research}.
\newblock DOI 10.3386/w30886.

\bibitem[Vincent, 2023a]{Vincent2023legal}
Vincent, J. (2023a).
\newblock Getty images is suing the creators of ai art tool stable diffusion
  for scraping its content.
\newblock Retrieved from
  https://www.theverge.com/2023/1/17/23558516/ai-art-copyright-stable-diffusion-getty-images-lawsuit.
\newblock Accessed on February 1, 2023.

\bibitem[Vincent, 2023b]{Vincent202382}
Vincent, J. (2023b).
\newblock Hustle bros are jumping on the ai bandwagon.
\newblock {\em The Verge}.

\bibitem[Volpicelli, 2023]{Volpicelli2023}
Volpicelli, G. (2023).
\newblock Chatgpt broke the eu plan to regulate {A}{I}. {P}olitico.
\newblock Retrieved from
  https://www.politico.eu/article/eu-plan-regulate-chatgpt-openai-artificial-intelligence-act/.
\newblock Accessed on March 22, 2023.

\bibitem[Warzel, 2023a]{Warzel2023118}
Warzel, C. (2023a).
\newblock Talking to {A}{I} might be the most important skill of this century.
\newblock Retrieved from
  https://www.theatlantic.com/technology/archive/2023/02/openai-text-models-google-search-engine-bard-chatbot-chatgpt-prompt-writing/672991/.
\newblock Accessed on March 21, 2023.

\bibitem[Warzel, 2023b]{WarzelVertigo2023}
Warzel, C. (2023b).
\newblock Welcome to the era of {A}{I} vertigo.
\newblock Retrieved from
  https://www.theatlantic.com/technology/archive/2023/02/google-bing-race-to-launch-ai-chatbot-powered-search-engines/673006/.
\newblock Accessed on February 26, 2023.

\bibitem[Weise and Metz, 2023]{Weise2023144}
Weise, K. and Metz, C. (2023).
\newblock Microsoft considers more limits for its new a.i. chatbot.
\newblock Retrieved from
  https://www.nytimes.com/2023/02/16/technology/microsoft-bing-chatbot-limits.html.
\newblock Accessed on February 26, 2023.

\bibitem[Weizenbaum, 1966]{Weizenbaum1966}
Weizenbaum, J. (1966).
\newblock Eliza a computer program for the study of natural language
  communication between man and machine.
\newblock {\em Communications of the ACM}, 9:36--45.

\bibitem[Wiggers, 2023]{Wiggers202380}
Wiggers, K. (2023).
\newblock The current legal cases against generative {A}{I} are just the
  beginning.
\newblock Retrieved from
  https://techcrunch.com/2023/01/27/the-current-legal-cases-against-generative-ai-are-just-the-beginning/?guccounter=1.
\newblock Accessed on February 7, 2023.

\bibitem[Williams, 2023]{Williams202353}
Williams, N. (2023).
\newblock Chatgpt and defining humanity in the age of brains in vats.
\newblock {\em Church Life Journal | University of Notre Dame}.
\newblock Accessed on March 21, 2023.

\bibitem[Wong, 2023]{Wong2023gpt4}
Wong, M. (2023).
\newblock Gpt-4 is here. just how powerful is it?
\newblock Retrieved from
  https://www.theatlantic.com/technology/archive/2023/03/gpt4-release-rumors-hype-future-iterations/673396/.
\newblock Accessed on March 19, 2023.

\end{thebibliography}

\end{document}